\documentclass{ieeeaccess}

\newcommand{\hl}{\textcolor{blue}}

\newcommand{\hlbegin}{\color{blue}}
\newcommand{\hlend}{\color{black}}
\renewcommand{\hl}{}

\renewcommand{\hlbegin}{}
\renewcommand{\hlend}{}

\newcommand{\hlx}{\textcolor{blue}}
\newcommand{\hlmx}{\mathcolor{blue}}
\renewcommand{\hlx}{}
\renewcommand{\hlmx}{}

\def\parameter{\bgroup\colorlet{outcolor}{.}\color{red}Y\futurelet\next\parameterA}
\def\parameterA{\ifx\next_\expandafter\parameterB\else\egroup\fi}
\def\parameterB_#1{_{{\color{outcolor}#1}}\futurelet\next\parameterC}
\def\parameterC{\ifx\next^\expandafter\parameterD\else\egroup\fi}
\def\parameterD^#1{^{\color{outcolor}#1}\egroup}

\usepackage{cite}
\usepackage{amsmath,amssymb,amsfonts}
\usepackage{mathtools}
\usepackage{algorithmic}
\usepackage{graphicx}
\usepackage{caption}
\usepackage{textcomp}
\usepackage{comment}
\usepackage{tabularx}
\usepackage{enumitem}
\newlist{tabitemize}{itemize}{1}
\setlist[tabitemize]{nosep,     
                     topsep     = 2pt       ,
                     leftmargin = *         ,
                     label      = $\bullet$ ,
                     }
\usepackage{multirow}

\setlist[itemize]{leftmargin=13pt}
\setlist[enumerate]{leftmargin=13pt}

\ifCLASSOPTIONcompsoc
    \usepackage[caption=false, font=normalsize, labelfont=sf, textfont=sf]{subfig}
\else
\usepackage[caption=false, font=footnotesize]{subfig}
\fi
\def\BibTeX{\mathrm{B\kern-.05em{\sc i\kern-.025em b}\kern-.08em
    T\kern-.1667em\lower.7ex\hbox{E}\kern-.125emX}}
\begin{document}
\history{Date of publication xxxx 00, 0000, date of current version xxxx 00, 0000.}
\doi{10.1109/TQE.2020.DOI}

\title{
Mitigating Precision Errors in Quantum Annealing via Coefficient Reduction of Embedded Hamiltonians
}
\author{
        \uppercase{Kentaro~Ohno}\authorrefmark{1},         \uppercase{Nozomu~Togawa}\authorrefmark{1},~\IEEEmembership{Member,~IEEE}
}
\address[1]{Department of Computer Science and Communications Engineering, Waseda University, Tokyo, Japan}


\markboth
{Author \headeretal: Preparation of Papers for IEEE Transactions on Quantum Engineering}
{Author \headeretal: Preparation of Papers for IEEE Transactions on Quantum Engineering}

\corresp{Corresponding author: Kentaro Ohno (email: kentaro.ohno@togawa.cs.waseda.ac.jp).}

\begin{abstract}
Quantum annealing is a quantum algorithm to solve combinatorial optimization problems.
In the current quantum annealing devices, the dynamic range of the input Ising Hamiltonian, \hl{defined as the ratio of the largest to the smallest coefficient,} \hl{significantly} affects the quality of the output solution due to limited hardware precision.
Several methods have been proposed to reduce the dynamic range by reducing large coefficients in the Ising Hamiltonian.
However, existing studies do not take into account minor-embedding, which is an essential process in current quantum annealers.
In this study, we revisit three existing coefficient-reduction methods \hlx{under the constraints of minor-embedding.} 
We evaluate to what extent these methods reduce the dynamic range of the minor-embedded Hamiltonian and improve the sample quality obtained from the D-Wave Advantage quantum annealer.
The results show that, on the set of problems tested in this study, the interaction-extension method effectively improves \hl{the sample quality} by reducing the dynamic range, while the bounded-coefficient integer encoding and the augmented Lagrangian method have only limited effects.
Furthermore, we empirically show that 
\hl{reducing external field coefficients at the logical Hamiltonian level is not required in practice},
since minor-embedding automatically has the role of reducing them.
These findings suggest future directions for enhancing the sample quality of quantum annealers \hlx{by suppressing hardware errors through} preprocessing of the input problem.
\end{abstract}

\begin{keywords}
Combinatorial optimization, Quantum annealing, Numerical precision, Minor-embedding
\end{keywords}

\titlepgskip=-15pt

\maketitle

\section{Introduction}
\label{sec:introduction}

\PARstart{Q}{uantum} annealing~\cite{apolloni1989quantum,somorjai1991novel,amara1993global,finnila1994quantum,kadowaki1998quantum} has been actively studied theoretically and experimentally for practical applications of quantum algorithms, since its physical realization by D-Wave systems Inc. in 2011~\cite{johnson2011quantum}.
\hlx{In the context of combinatorial optimization,} quantum annealing formulates a given optimization problem as an Ising Hamiltonian and searches for its ground state.
Theoretically, a wide class of combinatorial optimization problems can be modeled with the Ising Hamiltonian~\cite{lucas2014ising}. 
\hl{Consequently, quantum annealing is widely expected to serve as a powerful tool for solving various hard real-world problems~\cite{yarkoni2022quantum}.}

\hl{
Nevertheless, even with significant progress in hardware, 
current quantum annealers often fail to yield high-quality solutions for complex-structured problems (e.g., those with dense interactions or intricate constraints)~\cite{huang2022benchmarking,jiang2023classifying}, which is an obstacle to their practical use.
A primary factor contributing to this limited performance is the low numerical precision of the hardware~\cite{dickson2013thermally,young2013adiabatic,bian2014discrete,albash2015consistency,king2015performance,karimi2019practical,pearson2019analog,yarkoni2022quantum}.}
The external magnetic fields and couplings of the input Ising Hamiltonian are subject to various noises such as control errors and thermal noise, which possibly change the ground state of the Hamiltonian~\cite{albash2019analog}.
\hl{In particular}, an Ising Hamiltonian with a large dynamic range, i.e. the ratio between large and small coefficients in the Hamiltonian, is highly susceptible to such noise, resulting in poor-quality samples from the quantum annealer.

To address this problem, several techniques have been proposed to reduce the dynamic range by reducing large coefficients in the Hamiltonian.
Oku et al.~\cite{oku2020reduce} devised an interaction-extension method (IEM) that breaks down strong couplings into weak couplings while preserving the ground state by adding variables.
Karimi and Ronagh~\cite{karimi2019practical} proposed bounded-coefficient encoding (BCE) of integer variables, a method to control the maximum coefficient in linear combinations of binary variables representing integer variables.
Tanahashi and Tanaka~\cite{tanahashi2021augmented} suggested applying the augmented Lagrangian method (ALM) to quantum annealing to reduce the penalty coefficients for expressing constraints.

However, the analyses in the existing studies do not take minor-embedding~\cite{choi2008minor,choi2011minor} into account.
Minor-embedding is \hl{the process of mapping} the connectivity graph of an Ising Hamiltonian \hl{onto} the hardware graph of a quantum annealer.
\hl{This procedure} is essential \hl{for} practical applications since most real-world problems cannot be directly embedded as subgraphs of a sparse hardware graph.
\hl{At the same time, the embedding process also alters} the dynamic range of the Ising Hamiltonian.
Therefore, it is important to study the effect of the existing coefficient-reduction methods on the minor-embedded Hamiltonian to validate their practical utility.

In this study, we revisit the existing coefficient-reduction approaches under minor-embedding. 
We closely track how the coefficients change during minor-embedding, dividing it into the effect on the external field terms and that on the coupling terms.
In particular, the most significant coefficient in the minor-embedded Hamiltonian is typically given by the \emph{chain strength}, a parameter in minor-embedding that critically affects the \hl{sample quality} of quantum annealers.
\hl{To rigorously evaluate the reduction effect on the embedded Hamiltonian, it is essential to precisely identify the optimal chain strength \hlx{in terms of sample quality, e.g., average energy of samples or the rate of optimal solutions}.
Although existing heuristics~\cite{raymond2020improving,gilbert2024quantum} can provide estimates, they do not always yield the true optimum due to the problem-dependent nature of the energy landscape.
Therefore, we determine the optimal chain strength through an exhaustive search using actual hardware.}
This is in sharp contrast with the \hl{analyses} without minor-embedding, where the reduction effect can be assessed by theory or numerical simulation~\cite{karimi2019practical,oku2020reduce}, and is a major technical contribution of this paper.

We verify the effects of the three coefficient-reduction methods, the IEM, BCE, and ALM, on the minor-embedded Hamiltonian through extensive experiments using the D-Wave Advantage~\cite{Dwaveadvantage} quantum annealer.
The methods are tested on the \hl{instances from well-established benchmark datasets for}
the quadratic unconstrained binary optimization (QUBO) problem, the multi-dimensional knapsack problem (MKP), and the quadratic assignment problem (QAP).
The results show that while the IEM successfully reduces large coefficients even for minor-embedded Hamiltonians and improves the \hl{sample quality}, the effects of the BCE and ALM are of limited practical utility at least on the tested problems.
Furthermore, we found that \hlx{the coefficient reduction for the external field is not necessarily required in practice, since the external field coefficients are automatically reduced by minor-embedding}. 
These findings would be useful in guiding future directions of research and development for improving the performance of quantum annealing by suppressing \hlx{errors} in actual machines.
\hlx{Notably, our evaluation methodology is applicable regardless of whether a reduction method preserves the ground state, although the methods tested here are all ground-state preserving. This versatility allows our framework to quantify the practical utility of non-ground-state-preserving heuristics that may be developed in the future. Such assessments would offer new strategies for enhancing quantum annealing performance by exploring a broader range of coefficient-reduction approaches.}

The rest of this paper is organized as follows.
Section~\ref{sec:preliminary} provides the background of this study.
Section~\ref{sec:existing_methods} reviews the existing coefficient-reduction methods for quantum annealing.
In Section~\ref{sec:coefficient_change_embedding}, we discuss the expected effect of the methods on minor-embedded Hamiltonian.
The experimental results are presented in Section~\ref{sec:experiments}.
Section~\ref{sec:conclusion} concludes this paper with the discussion of future directions.

\section{Quantum Annealers}\label{sec:preliminary}

Quantum annealing is a quantum algorithm to heuristically solve combinatorial optimization problems by encoding them into ground states of a quantum Hamiltonian, proposed as an analogue of simulated annealing~\cite{apolloni1989quantum,somorjai1991novel,amara1993global,finnila1994quantum,kadowaki1998quantum}.
Quantum annealers manufactured by D-Wave Systems Inc. realize quantum annealing with superconducting flux qubits~\cite{johnson2011quantum} to find the ground state of the Ising Hamiltonian
\begin{align}\label{eq:hamiltonian}
    H = \sum_{i,j} J_{ij}\sigma_i \sigma_j + \sum_i h_i \sigma_i,
\end{align}
where $\sigma_i$ are spin variables taking either $+1$ or $-1$, represented by qubits, and the external fields $h_i$ and couplings $J_{ij}$ are given as weights encoding
a problem to solve.
An equivalent formulation known as unconstrained binary quadratic programming (UBQP)~\cite{kochenberger2014unconstrained}, also called quadratic unconstrained binary optimization (QUBO)~\cite{punnen2022quadratic}, is useful \hlx{for modeling various}
practical problems~\cite{lucas2014ising}.
QUBO is defined as 
\begin{align}
    &\mathrm{Minimize\ } \sum_{i=1}^n \sum_{j=i}^n Q_{ij} x_i x_j \notag\\
    &\mathrm{subject\ to\ } x_i \in \{0,1\} \quad \mathrm{for\ } i=1,\ldots,n,
\end{align}
where $Q_{ij} \in \mathbb R$ represents the problem data.
QUBO is translated into the Ising Hamiltonian by substituting $x_i = (\sigma_i+1)/2$.
We refer to the objective value as \emph{energy}, as in the case of Ising Hamiltonian.
Current state-of-the-art quantum annealers have more than 5,000 qubits, each admitting 15 couplings~\cite{Dwaveadvantage}.
Realizing quantum annealing for arbitrary Hamiltonians with current devices involves two major limitations: numerical precision and graph topology of the Hamiltonian, which are described below.

\subsection{Numerical Precision in Quantum Annealers}

Actual quantum annealers accept as input only Ising Hamiltonians whose weights fall within \hl{specified ranges}.
For example, the acceptable ranges of $h_i$ and $J_{ij}$ for current D-Wave devices are defined as $[-4,4]$ and $[-2,1]$, respectively~\cite{Dwaveadvantage}.
The input Hamiltonian is rescaled to fit the ranges if it has larger coefficients.
Generally, given the acceptable ranges $[h^-, h^+]$ and $[J^-, J^+]$ of the external field and coupling with $h^-,J^- < 0$ and $h^+, J^+ > 0$,
we define the following quantities for the Hamiltonian $H$:
\begin{align}
    s_h &\coloneqq \max \left( \frac{\max_i h_i}{h^+}, \frac{\min_i h_i}{h^-} \right), \\
    s_J &\coloneqq \max \left( \frac{\max_{i,j} J_{ij}}{J^+}, \frac{\min_{i,j} J_{ij}}{J^-} \right), \\
    s_H &\coloneqq \max(s_h, s_J).
\end{align}
We call them \emph{scaling factors} of the external field, coupling, and Hamiltonian, respectively.
The Hamiltonian $H$ is divided by $s_H$ before input into the quantum annealer\footnote{More precisely, there are additional limits on the total coupling per qubit, which is omitted here for simplicity. 
See also D-Wave's documentation:
https://docs.dwavequantum.com/en/latest/quantum\_research/errors.html.},
\hl{resulting in the rescaled Hamiltonian
\begin{align}
    H' \coloneqq H/s_H. 
\end{align}
}This rescaling often produces terms with small coefficients that are susceptible to noises such as control errors and thermal excitation~\cite{dickson2013thermally,young2013adiabatic,bian2014discrete,albash2015consistency,king2015performance,karimi2019practical,albash2019analog,pearson2019analog,yarkoni2022quantum}.
Specifically, the rescaled input Hamiltonian \hl{$H'$} is perturbed by the noise 
$\delta H$, 
resulting in the ``wrong'' Hamiltonian~\cite{young2013adiabatic,pearson2019analog}
\hl{\begin{align}
    H'' 
    &= H' + \delta H \notag \\
    &= \sum_{i,j} \left(\frac{J_{ij}}{s_H} + \delta J_{ij} \right) \sigma_i \sigma_j + \sum_i \left(\frac{h_i}{s_H} + \delta h_i \right) \sigma_i.
\end{align}
}The magnitude of the noise terms $\delta h_i, \delta J_{ij}$ is roughly estimated to be 1-5\% on the D-Wave devices~\cite{albash2015consistency,king2015performance,karimi2019practical,yarkoni2022quantum}.
If the input Hamiltonian has a large dynamic range, i.e., $s_H / \min_i |h_i|$ or $s_H / \min_{ij} |J_{ij}|$ where the minimum is taken over non-zero weights, the noise $\delta H$ easily changes the ground state and degrades the \hl{sample quality} obtained from quantum annealers~\cite{albash2019analog}.

\hl{A possible} way to suppress the errors is quantum annealing correction (QAC)~\cite{pudenz2014error,young2013adiabatic,vinci2015quantum,pearson2019analog}, which utilizes error correction code in quantum annealing.
While QAC is empirically effective for improving sample quality, 
the number of qubits is multiplied by a \hl{factor that scales with} the noise resilience level.
The huge increase in the number of qubits severely restricts the input problem size.
In this paper, \hl{as another approach distinct from QAC,} we discuss reducing the dynamic range of the Hamiltonian by decreasing large coefficients of the external fields and couplings.

\subsection{Minor-Embedding}\label{subsec:minor-embedding}

\begin{figure}[t]
    \centering
    \includegraphics[width=0.96\linewidth]{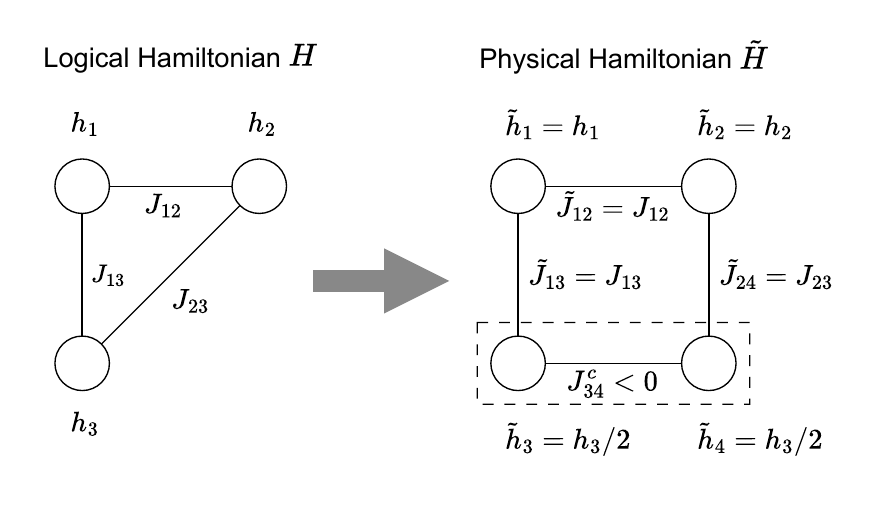}
    \caption{Example of minor-embedding of triangle graph onto square lattice. \hl{The chain $C(3) = \{3,4\}$ of the logical node $3$ is represented by the dashed box.}}
    \label{fig:example_minor_emnbedding}
\end{figure}

For the Ising Hamiltonian Eq.~(\ref{eq:hamiltonian}), an associated graph is defined where nodes and edges correspond to the variables $\sigma_i$ and non-zero couplings $J_{ij} \ne 0$, respectively. 
A major restriction of the quantum annealers is that they \hl{handle} 
Hamiltonians \hl{within} a fixed graph topology. 
\hl{In practice, a \emph{logical} Hamiltonian that represents the problem to be solved typically exhibits dense connectivity, which may not directly fit the hardware graph.}
To encode the logical Hamiltonian with general \hl{connectivity}, a technique called minor-embedding~\cite{choi2008minor,choi2011minor,cai2014practical} is used.
In graph theory, a \emph{minor} of a graph is defined as a subgraph of a graph obtained by contracting several edges in the original graph.
Minor-embedding transforms the logical Hamiltonian $H$ into a \emph{physical} Hamiltonian $\tilde H$ so that its associated graph is a subgraph of the hardware graph and edge contraction recovers the original problem graph.
An example of minor-embedding is shown in Fig.~\ref{fig:example_minor_emnbedding}.
A set of nodes to be merged via the edge contraction is called a \emph{chain}.
The coupling weights in $\tilde H$ consist of \hl{inter-chain weights}, 
which are inherited from the original Hamiltonian $H$, and \hl{intra-chain weights} \hl{intended to ensure the equivalence of $H$ and $\tilde H$ in terms of their ground states}.
More specifically, $\tilde H$ can be written as
\begin{align}
    \tilde H =& 
        \sum_{i} \sum_{k \in C(i)} \tilde{h}_k \sigma_k 
        + \sum_{i,j} \sum_{k \in C(i)} \sum_{l \in C(j)} \tilde{J}_{kl} \sigma_k \sigma_l 
        \notag\\ 
        &+ \sum_i \sum_{k,l\in C(i)} J^\mathrm{c}_{kl} \sigma_k \sigma_l,
\end{align}
where $C(i)$ is the chain for node $i$ in the original problem graph.
Couplings $\tilde{J}_{kl}$ and $J^\mathrm{c}_{kl}$ are zero if nodes $k$ and $l$ are not connected in the hardware graph.
The physical external fields $\tilde{h}_k$ and \hl{inter-chain weights} $\tilde{J}_{kl}$ are defined so that the following conditions hold:
\begin{align}\label{eq:embedded_h}
    \sum_{k\in C(i)} \tilde h_k &= h_i, \\
    \label{eq:embedded_J}
    \sum_{k \in C(i)} \sum_{l \in C(j)} \tilde{J}_{kl} &= J_{ij}.
\end{align}
The \hl{intra-chain weights} $J^\mathrm{c}_{kl}$ are set to negative numbers to encourage variables in a chain to take the same value.
The absolute value $|J^\mathrm{c}_{kl}|$ is called \emph{chain strength}.
The chain strength should be set sufficiently large to suppress \emph{chain break}, i.e., a sample from quantum annealers having inconsistent value assignment for variables in a chain.
\hl{On the other hand, it should not be so large that, after rescaling, other coefficients fall below the precision limit of hardware.}
\hlx{It is crucial to identify the optimal chain strength that maximizes sample quality for precise evaluation of quantum annealing performance.}
In this study, the chain strength takes the same value for every chain for simplicity. 


\section{Coefficient Reduction Methods}\label{sec:existing_methods}

We briefly review three existing approaches~\cite{oku2020reduce,karimi2019practical,tanahashi2021augmented} for reducing large coefficients in the Ising Hamiltonian.
All of them were discussed \hl{at} the logical Hamiltonian level, and their effects under minor-embedding have not been explored.
They have different applicability, advantages and disadvantages from each other, which are summarized in Table~\ref{tab:coefficient_reduction_method_summary}.
To implement and verify the approaches efficiently, a refined version of each method is introduced below.

\begin{table}
\centering
\caption{Existing coefficient-reduction methods.}
\label{tab:coefficient_reduction_method_summary}
\renewcommand{\tabularxcolumn}{m} 
\begin{tabularx}{\linewidth}{|c|>{\raggedright}X|}
\hline
Method & Properties \tabularnewline
\hline
IEM~\cite{oku2020reduce} &
\begin{tabitemize}
    \item Applicable to arbitrary Ising Hamiltonians.
    \item Additional variables for each large coupling. 
\end{tabitemize}
\tabularnewline
\hline
BCE~\cite{karimi2019practical} & 
\begin{tabitemize}
    \item Applicable only for representing integer variables.
    \item Additional variables for each integer variable.
\end{tabitemize}
\tabularnewline
\hline
ALM~\cite{tanahashi2021augmented} & 
\begin{tabitemize}
    \item Applicable only to penalty coefficients.
    \item No additional variables.
    \item Reduction rate is bounded and relatively small.
\end{tabitemize}
\tabularnewline
\hline
\end{tabularx}
\end{table}

\subsection{Interaction-extension Method}\label{subsec:interaction-extention}

\begin{figure}[t]
    \centering
    \includegraphics[width=0.96\linewidth]{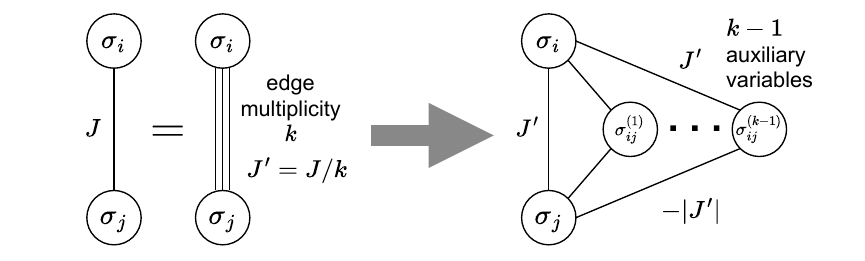}
    \caption{Procedure \hl{of} interaction-extension method~\cite{oku2020reduce}.}
    \label{fig:example_interaction_extention}
\end{figure}

Oku et al.~\cite{oku2020reduce} proposed a method to break a large coefficient down into smaller coefficients based on interaction-extension operations using auxiliary variables.
We call it the interaction-extension method (IEM).
Here, we introduce a version which is efficient in terms of the number of auxiliary variables, see also modification by Kikuchi et al.~\cite{kikuchi2023dynamical}.
We consider the Hamiltonian $H$ given in Eq.~(\ref{eq:hamiltonian}).
Suppose \hl{the goal is} to bound the maximum value of the couplings  to $M>0$.
The procedure is as follows:
for each positive coupling coefficient $J_{ij} > 0$ exceeding $M$, we define
\begin{align}
    \hl{k_{ij}} \coloneqq \left\lceil \frac{J_{ij}}{M} \right\rceil
\end{align}
and replace the term $J_{ij}\sigma_i \sigma_j$ in $H$ with
\begin{align}
    \frac{J_{ij}}{\hl{k_{ij}}}\sigma_i\sigma_j + \sum_{l=1}^{\hl{k_{ij}}-1} \frac{J_{ij}}{\hl{k_{ij}}}\sigma_i\sigma_{ij}^{(l)} - \sum_{l=1}^{\hl{k_{ij}}-1} \frac{J_{ij}}{\hl{k_{ij}}}\sigma_j\sigma_{ij}^{(l)},
\end{align}
where $\sigma_{ij}^{(l)}, l=1,2,\ldots,\hl{k_{ij}}-1$ are auxiliary spin variables.
Formally, the procedure is given by partitioning the original coupling $J_{ij}$ into \hlx{multiple edges} of weight $J_{ij}/\hl{k_{ij}}$ followed by the extension of those edges using additional nodes except one, as shown in Fig.~\ref{fig:example_interaction_extention}.
By \hl{construction}, all coupling weights in the resulting Hamiltonian do not exceed $M$.
In addition, it has the same ground state as the original Hamiltonian $H$, ignoring assigned values for the auxiliary spins, cf.~\cite[Theorem~2]{oku2020reduce}.
The negative couplings can be bounded from below by adding variables in a similar manner.
Reducing the range of the external fields is also possible, see the original paper~\cite{oku2020reduce}.
In this way, the method enables us to reduce the dynamic range of the Hamiltonian.
This method can be applied to arbitrary Ising Hamiltonians at the expense of the increase in the number of variables.

\subsection{Bounded-coefficient Integer Encoding}\label{subsec:bce}

Combinatorial optimization problems often involve integer variables.
In the context of quantum annealers, an integer variable is typically represented by a linear combination of multiple binary variables.
Namely, an integer variable $z$ taking a value in some range is represented as
\begin{align}\label{eq:integer_encoding}
    z = c + \sum_{i=1}^n \frac{a_i}{2} \sigma_i,
\end{align}
where $c$ and $a_i$ are constants and $\sigma_i$ are spin variables.
There are various choices of the linear combination, each offering different properties\hl{~\cite{ohno2024toward}}.
For example, binary \hlx{encoding}, which adopts $a_i = 2^{i-1}$, has the most compact representation in terms of the number of variables, while the maximum coefficient can be exponentially large.
Karimi and Ronagh proposed bounded-coefficient encoding (BCE)~\cite{karimi2019practical} to control the trade-off between the number of variables and the coefficient magnitude, 
introducing an upper bound parameter $\mu$ for the coefficients.
\hl{In this encoding, the coefficients $a_i$ are set so as not to exceed 
$\mu$ while keeping the number of binary variables as small as possible.
Specifically, this is achieved by using the binary \hlx{encoding} 
up to a certain point and then capping the coefficients at $\mu$ when they would otherwise exceed this limit. 
A formal description is given below.}

Let $z$ be an integer variable taking a value in a range $[L, U]$ for integers $L,U\in \mathbb Z$.
Define $D \coloneqq U- L$.
For a given coefficient bound $\mu \in [1,D]$, we define
\begin{align}\label{eq:bce_notation}
    m \coloneqq \left\lfloor D/\mu \right\rfloor, \
    r \coloneqq \mu + (D - \mu m),\
    k \coloneqq \lfloor \log_2 r \rfloor.
\end{align}
BCE for $z$ is defined as follows, using $k+m$ binary variables:
\begin{align}
    z &= L + \frac{D}{2} + \sum_{i=0}^{k+m-1} \frac{a_i}{2} \sigma_i,  \\ \label{eq:bce_coefficient}
    a_i &= \begin{cases}
        2^{i}   & \mathrm{for \ } i = 0, \ldots, k-1 \\
        r+1-2^k & \mathrm{for \ } i = k \\
        \mu     & \mathrm{for \ } i = k+1, \ldots, k+m-1.
    \end{cases}
\end{align}
\hl{The coefficients $a_i$ for $i \le k$ correspond to the binary expansion of $r$.
By construction, the condition $\sum_i a_i = D$ holds.
These properties ensure that this expansion yields a valid encoding of $z$. 
Namely, every integer in $[L,U]$ can be realized by varying the configurations of $\sigma_i$.
As a special case, setting $\mu = D$ results in $r=D$, which recovers the binary \hlx{encoding} of $z$.
Also note that the \hlx{encoding} can be expressed as $z = L + \sum_i a_i y_i$ using binary variables $y_i \in \{0, 1\}$, where each $y_i$ relates to the spin variable $\sigma_i$ as $y_i = (\sigma_i + 1)/2$. 
Since $r$ satisfies $r <2\mu$, $a_i$ do not exceed $\mu$ for $i\le k$.
For the remaining indices, $a_i$ are capped at the upper bound parameter $\mu$.
Therefore, $0 < a_i \le \mu$ holds for all $i=0,\ldots,k+m-1$.}

\begin{figure}[t]
    \centering
    \includegraphics[width=0.96\linewidth]{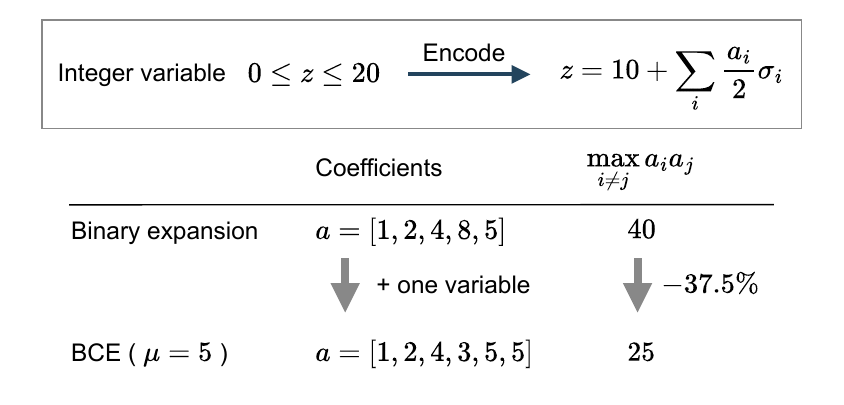}
    \caption{Example of bounded-coefficient encoding~\cite{karimi2019practical}.
    The coefficient-reduction effect on logical couplings is assessed through $\max_{i\ne j} a_i a_j$, assuming $z^2$ appears in the objective function.}
    \label{fig:example_BCE}
\end{figure}

\hl{Based on this formulation, the dynamic range of the Hamiltonian can be controlled by adjusting the upper bound parameter $\mu$.
Decreasing $\mu$ results in small coefficients, while it increases the number of variables, as illustrated in Fig.~\ref{fig:example_BCE}.
Although the notation in Eq.~(\ref{eq:bce_notation}) and Eq.~(\ref{eq:bce_coefficient}) differs from the original formulation~\cite{karimi2019practical}, both provide the exact same representation of $z$ up to ordering of indices.}

The BCE is particularly effective when a product or square of integer variables appears in the objective function.
For example, an integer variable is often used to represent a linear inequality constraint condition
\begin{align}
    L \le \sum_i b_i \hl{x_i} \le U,
\end{align}
where $b_i, L$ and $U$ are integers and \hl{$x_i$ are binary variables}. 
This constraint is translated into a penalty term using an integer variable $z$ as
\begin{align}
    \left(\sum_i b_i \hl{x_i} - z\right)^2 , \quad L\le z \le U.
\end{align}
The penalty produces a term $z^2$, which may include a large number of large coefficients when the \hlx{encoding} Eq.~(\ref{eq:integer_encoding}) involves large coefficients.
In that situation, the IEM 
\hl{becomes prohibitive in terms of variable overhead, as the number of required auxiliary variables scales with the magnitude and density of strong couplings.}
The BCE enables us to eliminate those large coefficients all at once, introducing a small number of additional variables.

\subsection{Augmented Lagrangian Method}\label{subsec:augmented_lagrangian}

Another major source of large coefficients in the Hamiltonian is penalty coefficients for representing constraint conditions.
A linear equality constraint condition
\begin{align}
    \sum_{i} b_i \sigma_i = c
\end{align}
on spin variables $\sigma_i$ is translated into a penalty term 
\begin{align}
    H_\mathrm{pen} = \left(\sum_{i} b_i \sigma_i - c\right)^2 = g(\sigma)^2.
\end{align}
Here, we defined $g(\sigma) \coloneqq \hlmx{\sum_{i} b_i \sigma_i - c}$.
Then, the penalty term is added to the objective function $H_\mathrm{obj}$ with positive scaling to obtain a problem Hamiltonian
\begin{align}
    H = H_\mathrm{obj} + \lambda H_\mathrm{pen}.
\end{align}
The penalty coefficient $\lambda>0$ should be set sufficiently large to obtain feasible solutions.
Appropriate values for $\lambda$ can be quite large depending on the problem.
For example, it can be around $10^4$ even for a small-scale quadratic assignment problem (QAP) instance, see Section~\ref{subsec:experiment_augmented_lagrangian}.
Note that multiple constraint conditions may be imposed in general, but here we consider the single constraint case for simplicity, as the extension to multiple constraints is straightforward.

\hlx{In continuous optimization, }\hl{the penalty method 
\hlx{often causes} numerical issues, such as ill-conditioning, particularly with large penalty coefficients.}
To reduce the penalty coefficient, a method now called the augmented Lagrangian method (ALM) was developed~\cite{hestenes1969multiplier}.
In the context of quantum annealing, Tanahashi and Tanaka proposed applying the ALM also for QUBO formulation~\cite{tanahashi2021augmented}.
The ALM involves minimization of the \emph{augmented Lagrangian function}
\begin{align}\label{eq:augmented_lagrangian}
    L = H_\mathrm{obj} + u g(\sigma) + \lambda g(\sigma)^2.
\end{align}
Here, $u \in \mathbb R$ is the Lagrange multiplier for the constraint.
As in the continuous optimization case, tuning of $u$ and $\lambda$ is based on iterative updates using a solution $\sigma^*$ for fixed $u, \lambda$:
\begin{align}\label{eq:alm_update}
    u \leftarrow u + 2 \lambda g(\sigma^*), \quad
    \lambda \leftarrow  \alpha \lambda,
\end{align}
where $\alpha>1$ is a constant.
Although the ALM is expected to reduce the penalty coefficient, \hlx{its reduction effect for QUBO has not been quantitatively evaluated in the existing study~\cite{tanahashi2021augmented}.
This is because the optimal ranges for the parameters $u$ and $\lambda$ are difficult to pre-determine and their interplay is non-intuitive, which forces the method to follow a specific, simultaneous update path as in Eq.~(\ref{eq:alm_update}). Since this path-dependent approach does not explore the full parameter space, it may not yield the minimum possible coefficients.}


\hlx{To address this limitation}, we introduce another interpretation of the ALM formulation for QUBO. 
The augmented Lagrangian function Eq.~(\ref{eq:augmented_lagrangian}) can be rewritten as 
\begin{align}
    L = H_\mathrm{obj} + \lambda \left(g(\sigma) - \epsilon \right)^2 - \lambda \epsilon^2,
\end{align}
with $\epsilon \coloneqq - u / 2\lambda$.
The last term is constant, thus can be ignored for optimization.
The second term can be seen as a penalty term that corresponds to a \emph{perturbed} constraint $g(\sigma)=\epsilon$ reformulated from the original constraint $g(\sigma)=0$.
If the perturbation width $|\epsilon|$ is large, the constraint is not valid anymore.
On the other hand, if $|\epsilon|$ is \hl{sufficiently} small, $\sigma$ minimizing $(g(\sigma)-\epsilon)^2$ also minimizes $g(\sigma)^2$ since the variable $\sigma$ is discrete.
Proper perturbation possibly increases the penalty without increasing the penalty coefficient $\lambda$.
For instance, consider a one-hot constraint
\begin{align}
    \sum_{i=1}^n x_i = 1
\end{align}
on binary variables $x_i \in \{0, 1\}, i=1,\ldots, n$,
imposing $x_i=1$ for exactly one index $i$.
Note that $x_i$ relate to spin variables $\sigma_i$ as $x_i = (\sigma_i+1)/2$.
For $g(x)=\sum_{i=1}^n x_i - 1$, 
\hlx{the perturbed penalty for an infeasible solution $\mathbf 0=(0,\ldots,0)$ is given as
\begin{align}
    (g(\mathbf{0}) - \epsilon)^2 - \epsilon^2 = 1 + 2\epsilon.
\end{align}
Therefore, in cases where small $\lambda$ values lead to the infeasible solution $x=\mathbf{0}$ under the usual penalty method, a positive perturbation $\epsilon>0$ 
effectively imposes a larger penalty. 
Such a situation arises, e.g., in the QAP (Section~\ref{sec:experiments}).
The  valid perturbation width is given as $|\epsilon|<0.5$, see Table~\ref{tab:example_penalty_perturbation}.
By setting $\epsilon$ close to $0.5$, the penalty for $x=\mathbf 0$ increases by up to a factor of 2.
Thus, the penalty coefficient $\lambda$ is reduced by at most 50\%.
These properties enable a transparent and systematic evaluation of the coefficient-reduction effect of the ALM, as demonstrated through experiments on the QAP in Section~\ref{sec:experiments}.}

\begin{table}[t]
    \centering
    \caption{Example of perturbation of one-hot constraint.}
    \label{tab:example_penalty_perturbation}
    \begin{tabular}{c|ccc}
            &             & Original & Perturbed Penalty \\
        $x$ & Feasibility & $\left( \sum_i x_i - 1 \right)^2 $ & $\left( \sum_i x_i - 1 - \epsilon \right)^2 - \epsilon^2$ \\
        \hline
        $[0,0,0]$ & No  & $1$ & $1+2\epsilon$ \\
        $[1,0,0]$ & Yes & $0$ & $0$ \\
        $[1,1,0]$ & No  & $1$ & $1-2\epsilon$
    \end{tabular}
\end{table}

\section{Coefficients in Embedded Hamiltonian}\label{sec:coefficient_change_embedding}

Minor-embedding \hl{alters} the coefficients in the Hamiltonian as shown in Section~\ref{subsec:minor-embedding}.
In this section, we discuss how it relates to the coefficient reduction of
the logical Hamiltonian.

First, we recall how coefficients in the Hamiltonian change through minor-embedding.
Among possible \hl{physical external fields $\tilde{h}_k$ and inter-chain weights $\tilde{J}_{kl}$} satisfying Eq.~(\ref{eq:embedded_h}) and Eq.~(\ref{eq:embedded_J}), it is typical to set them to be balanced:
\begin{align}\label{eq:embedded_h_2}
    \tilde{h}_k &\coloneqq \frac{h_i}{|C(i)|} \quad \mathrm{for\ } k \in C(i),\\
    \tilde{J}_{kl} &\coloneqq \frac{J_{ij}}{|S(i,j)|} \quad \mathrm{for\ } (k,l) \in S(i,j),
\end{align}
where $S(i,j) \subset C(i)\times C(j)$ is a set of pairs of nodes connected in the hardware graph.
Practically, it happens only occasionally that $|S(i,j)| >1$ holds.
Furthermore, the \hl{inter-chain weights} are usually not dominant since the \hl{intra-chain weights} are typically larger.
Therefore, one may assume 
\hlx{$|S(i,j)|=1$}
for simplicity.
In any case, Eq.~(\ref{eq:embedded_h_2}) shows that minor-embedding has the effect of reducing large coefficients of the external field.
More precisely, it can make small coefficients even smaller, which rather increases the dynamic range.
To avoid this, we may set $\tilde{h}_k = h_i$ for exactly one $k \in C(i)$ for small $h_i$.
In practice, it can rarely be an issue, since $h_i$ tend to take large values compared with the couplings $J_{ij}$ on practical problems formulated in a QUBO form.
\hlx{Hence, we adopt Eq.~(\ref{eq:embedded_h_2}) to compute $\tilde{h}_k$ for simplicity}. 
The \hl{intra-chain weight}, i.e., chain strength, is generally a more significant factor \hl{that contributes to} the dynamic range of the embedded Hamiltonian.
Namely, the \hl{optimal} chain strength is usually larger than the absolute values of the \hl{inter-chain weights}, 
\hl{as observed in our experiments (Section~\ref{sec:experiments}) and supported by
previous studies~\cite{raymond2020improving,berwald2023understanding,djidjev2023logical,gilbert2024quantum}}.
In summary, minor-embedding reduces 
the external field coefficients
while 
raising the maximum 
coupling coefficients.

Now we revisit the existing coefficient-reduction methods, 
accounting for the effects of minor-embedding.
By reducing the coefficients of couplings $J_{ij}$ in the logical Hamiltonian, the methods directly reduce the \hl{inter-chain weights} $\tilde J_{kl}$ in the physical Hamiltonian.
The reduction also has an implicit effect on the \hl{intra-chain weights} $J_{kl}^c$, possibly decreasing the \hl{optimal} chain strength.
We expect that 
the reduction rate for the \hl{optimal} chain strength would be as high as that of the \hl{inter-chain weights} in an ideal case.
In this way, the existing methods on the logical Hamiltonian would also work to some extent on the physical Hamiltonian.
On the other hand, minor-embedding itself has the effect of reducing the range of the external field $\tilde{h}_k$ as shown in Eq.~(\ref{eq:embedded_h_2}).
Therefore, \hl{it is not necessary to} reduce the coefficients of the external fields $h_i$ in the logical Hamiltonian if the physical external fields $\tilde{h}_k$ are already \hl{sufficiently} small compared with the couplings.
This aspect is unique to our setting in contrast with the earlier work~\cite{karimi2019practical,oku2020reduce}, which considered reducing the external fields in the logical Hamiltonian as well.

\begin{table*}[t]
    \centering
    \caption{\hl{Overview of Experimental Designs and Problem Instances.}}
    \label{tab:experiment_overview}
    \begin{tabular}{l|lllll}
        \hline
        ID     & Section No. & Purpose  & Problem & Source  & Instances \\
        \hline
        \hline
        Exp.~1 & Section~\ref{subsec:h_reduction_by_minor-embedding} & Necessity of reducing external fields & Various domains & All of the below & 126 QUBO instances in MQLIB, etc. \\
        \hline
        \multirow{2}{*}{Exp.~2} & \multirow{2}{*}{Section~\ref{subsec:experiment_iem}} &\multirow{2}{*}{Validation of IEM} & Trivial Ising problem & Custom  & $\min_\sigma (J \sigma_1 \sigma_2 + \sigma_2 \sigma_3)$ \\
               &                   & & QUBO & MQLIB~\cite{DunningEtAl2018MQLIB} & gka1b, gka2b, gka3b \\
        \hline
        \multirow{2}{*}{Exp.~3} & \multirow{2}{*}{Section~\ref{subsec:experiment_bce}} & \multirow{2}{*}{Validation of BCE} & Trivial integer problem & Custom  & $\min_{z} (z-1)^2$ \\
               &                   & & MKP & OR-Library~\cite{beasley1990or} & weing1, weish06 \\
        \hline
        Exp.~4 & Section~\ref{subsec:experiment_augmented_lagrangian} & Validation of ALM & QAP & \cite{nugent1968experimental,taillard1991robust} & nug5, tai5a \\
        \hline
    \end{tabular}
\end{table*}

\hl{Given the above discussion, the primary experimental objectives} are summarized as follows:
\begin{itemize}
    \item \textbf{Necessity of reducing external fields}: We explore how often the physical external fields $\tilde h_{k}$ can be problematically large in practical situations.
    \item \textbf{Reduction effects on chain strength}: We investigate to what extent the coefficient reduction of the logical Hamiltonian decreases the \hl{optimal} chain strength.
    \item \textbf{Improvement of \hl{sample quality}}: We test whether the methods improve the overall \hl{sample quality} of quantum annealers. 
\end{itemize}
Regarding the first point, if $\tilde{h}_k$ tend to be much smaller than the couplings, it would suggest that the reduction of $h_i$ is of low importance in practice, as well as justify an approach to reduce the coefficient range of couplings at the risk of increasing $h_i$, such as the ALM.
The second point requires comprehensive tuning of chain strength based on sampling from actual quantum annealers. 
This is technically more involved compared with the analysis of the reduction effect on the logical Hamiltonian, where the effect can be assessed by theory or simulation~\cite{karimi2019practical,oku2020reduce}.
As for the third point, additional cost due to the methods, e.g., the increase in the number of variables, might change minor-embedding significantly, which could rather degrade the performance of quantum annealing.
These aspects are addressed through experimental evaluations in Section~\ref{sec:experiments}.

\section{Experiments}\label{sec:experiments}

\subsection{Experimental Setup}\label{subsec:setup}

\hlbegin


This section evaluates the coefficient-reduction methods through four experiments.
Each experiment is designed to test a specific aspect, as detailed in the following subsections:

\begin{itemize}
    \item \textbf{Exp. 1 (Analysis of Physical External Fields)}: Conducts a comprehensive analysis of 130 instances across multiple domains to evaluate the general necessity of reducing external fields. (Section~\ref{subsec:h_reduction_by_minor-embedding})
    \item \textbf{Exp. 2 (Verification of IEM)}: \hlx{Observes the impact of limited hardware precision on sample quality and how it is mitigated by the IEM on QUBO instances.}
    (Section~\ref{subsec:experiment_iem})
    \item \textbf{Exp. 3 (Verification of BCE)}: \hlx{Evaluates} 
    the performance of the BCE in suppressing coefficient growth during integer-to-binary encoding \hlx{using the multi-dimensional knapsack problem (MKP)}. (Section~\ref{subsec:experiment_bce})
    \item \textbf{Exp. 4 (Verification of ALM)}: Assesses the effectiveness of the ALM on the quadratic assignment problem (QAP) involving large penalty coefficients. (Section~\ref{subsec:experiment_augmented_lagrangian})
\end{itemize}

\subsubsection{Problem Datasets}

To evaluate the reduction methods under diverse conditions, the benchmark datasets in this study are derived from three distinct problem classes, all of which are formulated in the QUBO \hlx{or Ising} form for submission to quantum annealers.
These include random QUBO instances from MQLIB~\cite{DunningEtAl2018MQLIB}, MKP instances from OR-Library~\cite{beasley1990or}, and QAP instances from the literature~\cite{nugent1968experimental,taillard1991robust}.
As summarized in Table~\ref{tab:experiment_overview}, we additionally include custom-made trivial instances for Exp.~2 and 3 to provide a clear baseline for observing the fundamental behavior of the evaluated methods. 
The specific configurations and formulations for each evaluation are provided within their respective subsections. 

\subsubsection{Evaluation Workflow}
We employ the following
workflow consisting of three steps:

\begin{enumerate}
    \item \textbf{Minor-embedding}: Mapping the logical Hamiltonian onto the hardware graph using a heuristic algorithm.
    \item \textbf{Scaling factor evaluation}: Assessing the physical scaling factors for the embedded Hamiltonian.
    \item \textbf{Evaluation of optimal chain strength and sample quality}: Conducting a grid search for the chain strength to identify its optimal value by sampling solutions from the quantum annealer.
    The sample quality is then evaluated at the obtained optimal chain strength.
\end{enumerate}

For the comprehensive analysis in Exp.~1, we execute Step~1 and Step~2 for all 130 instances to focus on general scaling factor trends.
Additionally, for instances where the scaling factor of the physical external field cannot be judged as sufficiently small after Step~2, we also perform Step~3 to evaluate the scaling factor relative to the optimal chain strength.
In fact, only a few instances required Step~3; since these instances were included in Exp.~2, 
Exp.~1 incorporated the data from Exp.~2 to ensure consistency.

All three steps are performed in Exp.~2--4 to analyze the impact of each 
method.
Note that Step~3 primarily focuses on the effect of the methods on physical coupling coefficients and sample quality.
To justify this focus, it is essential to ensure that the hardware precision bottleneck is not shifted to the external field as the coupling coefficients are reduced.
Therefore, we monitor the scaling factor $s_{\tilde h}$ in Step~2 to confirm that the physical external field coefficients become sufficiently small relative to the couplings throughout Exp.~2--4, which consequently reinforces the findings of Exp.~1.

\hlx{
In Step~3, sample quality is measured by the probability of obtaining the ground state, denoted as $P_{\mathrm{opt}}$, when identifiable, or by the average energy for more complex problems. The optimal chain strength is determined based on the average energy, as $P_{\mathrm{opt}}$ is susceptible to high stochastic variance. For constrained problems, the objective value of feasible solutions is adopted as a practical utility metric.}

\hlend

\subsubsection{Computational Setup}

We use the D-Wave Advantage~\cite{Dwaveadvantage} quantum annealer.
The hardware graph is a Pegasus graph $P_{16}$ having more than 5,000 nodes.
The annealing time is set to 0.1 ms in our experiments if it is not specified.
We use D-Wave Ocean SDK\footnote{https://github.com/dwavesystems/dwave-ocean-sdk} version 8.1.0 to run the experiments.
The code is run on a MacBook Pro with Apple M2 chip and 8GB memory.
For minor-embedding search, we adopt clique-based minorminer~\cite{zbinden2020embedding}.
Specifically, we use the minorminer algorithm~\cite{cai2014practical} implemented in D-Wave Ocean SDK, setting the clique embedding~\cite{Dwaveadvantage} to the initial-embedding parameter.
Other parameters are set to default.
Chain breaks in samples from the quantum annealer are fixed by majority vote.

\subsection{Reduction of External Fields through Minor-embedding}\label{subsec:h_reduction_by_minor-embedding}

\hl{In this experiment, we investigate the necessity of reducing the external field coefficients of the logical Hamiltonian. Specifically,} we compute the physical external field and compare it with the physical coupling on various \hl{problems}.
For the scaling factors $s_{\tilde{h}}, s_{\tilde{J}}$ of the physical external field and coupling, an inequality $s_{\tilde{h}} \le s_{\tilde{J}}$ implies that coefficient reduction \hl{is unnecessary} for the external fields.
Note that $s_{\tilde{J}}$ depends not only on the \hl{inter-chain weights}, but also on the \hl{intra-chain weights}, i.e., chain strength.
Since the scaling factor $s_{J}$ of the logical couplings is generally smaller and computationally more tractable than $s_{\tilde{J}}$, we mainly 
\hl{verify a stronger inequality} $s_{\tilde{h}} \le s_J$ \hl{using} the benchmark instances.

\hl{We demonstrate the evaluation on} a collection of QUBO instances in a benchmark dataset called MQLIB~\cite{DunningEtAl2018MQLIB}.
The collection consists of 126 QUBO instances from existing studies~\cite{beasley1998heuristic,glover1998adaptive,palubeckis2006iterated}.
We exclude other instances in MQLIB since they are given as the max-cut problem, \hlx{which does not require external field terms}. 
We run minor-embedding search for the test instances with the target graph $P_{16}$.
For obtained valid embeddings, 
we compute the scaling factor $s_{\tilde h}$ of the physical external field.
For non-embeddable instances, we estimate $s_{\tilde h}$ assuming we have a sufficiently large Pegasus graph $P_m$ and embed the problem graph as a clique-minor.
The detail of the estimation is given in Appendix~\ref{app:physical_h_scaling_estimation}.

\begin{figure}[t]
    \centering
    \includegraphics[width=0.825\linewidth]{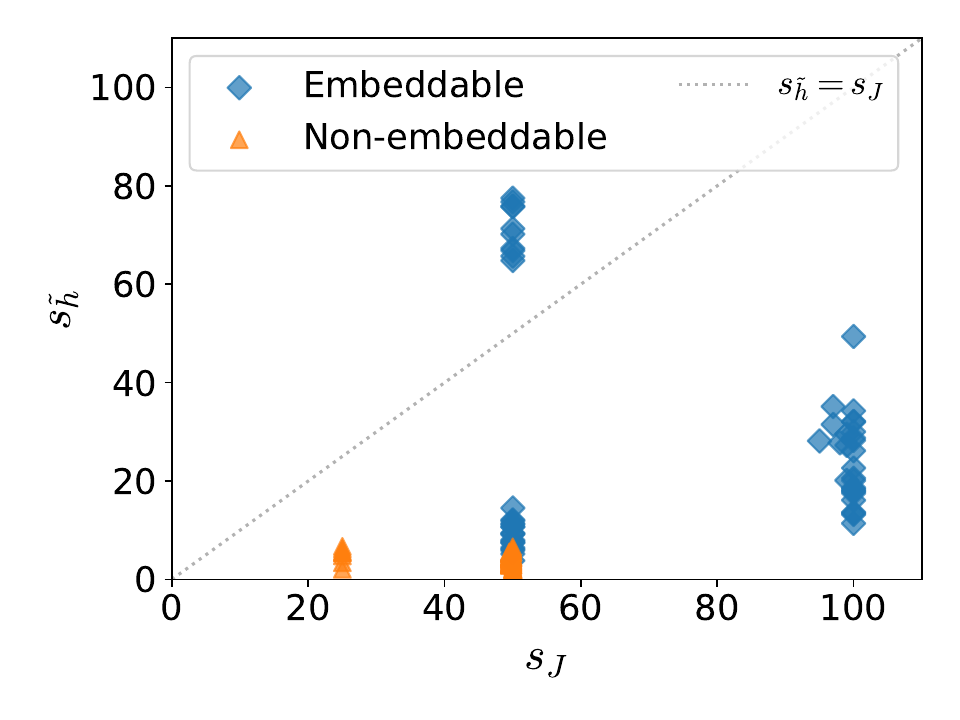}
    \caption{Scaling factors on QUBO instances in MQLIB.}
    \label{fig:mqlib_embedded_dr}
\end{figure}

\begin{table}[t]
    \centering
    \caption{Scaling factors on gka.*b instances.}
    \label{tab:gka_b_dr}
    \begin{tabular}{c|r|r|r}
    Instance & size & $s_{\tilde h} / s_J$ & $s_{\tilde J}/ s_{J}$  \\
    \hline
    gka.1b   & 20   & $1.31$   & $1.25 $    \\
    gka.2b   & 30   & $1.40$   & $2.00 $    \\
    gka.3b   & 40   & $1.30$   & $2.25 $    
    \end{tabular}
\end{table}

\hl{Fig.~\ref{fig:mqlib_embedded_dr} presents the comparison of $s_{\tilde h}$ with $s_J$ on the MQLIB instances.}
We observe that $s_{\tilde{h}} \le s_J$ holds on most of the instances.
There are ten exceptional instances, all of which are of the same problem type named gka.*b for * in $\{1,2,\ldots,10\}$.
\hl{For this problem class, we evaluate the magnitude of $\hlmx{s_{\tilde{J}}}$ more precisely \hlx{to verify} whether the condition $s_{\tilde{h}} \le s_{\tilde{J}}$ holds, where $s_{\tilde{J}}$ is calculated based on the optimal chain strength. To perform this comparison, we utilize the optimal chain strength data obtained from Exp.~2 in Section~\ref{subsec:experiment_iem}.}
Note that since the accepted range of couplings is $[-2, 1]$, $s_{\tilde J}$ is computed by dividing the \hl{optimal} chain strength by $2$ when the chain strength is larger than $2 s_{J}$.
The ratios $s_{\tilde h_k} / s_J$ and $s_{\tilde J}/ s_{J}$ on three instances, gka.1b, gka.2b, and gka.3b, are summarized in Table~\ref{tab:gka_b_dr}.
We observe that $s_{\tilde{h}} \le s_{\tilde{J}}$ holds except on gka.1b.
On gka.1b, the ratio $s_{\tilde{h}} / s_{\tilde{J}}$ is $1.31/1.25 \approx 1.05$, \hl{which appears to be \hlx{negligibly small} in terms of its impact on the sample quality.}

In summary, there are almost no cases where coefficient reduction for the external field is required in the MQLIB dataset.
\hl{Beyond the MQLIB instances originally defined as QUBO, we performed similar verifications for the MKP and QAP by examining whether the physical external field coefficients become sufficiently small after minor-embedding.
For the sake of consistency in notation and presentation, the underlying experimental data are provided in Section~\ref{subsec:experiment_bce} and Section~\ref{subsec:experiment_augmented_lagrangian}, respectively.
The results demonstrate that, consistent with the MQLIB instances, reduction of the external field in the logical Hamiltonian is not required for these problem instances.}
\hl{It was also confirmed that the inequality $s_{\tilde{h}} \le s_{\tilde{J}}$ consistently holds even when logical couplings $J_{ij}$ are reduced by the respective coefficient-reduction methods.
Experimental data supporting this observation are presented in Section~\ref{subsec:experiment_iem}, Section~\ref{subsec:experiment_bce}, and Section~\ref{subsec:experiment_augmented_lagrangian}.
}

\hlend


\begin{figure}[t]
    \centering
    \includegraphics[width=0.825\linewidth]{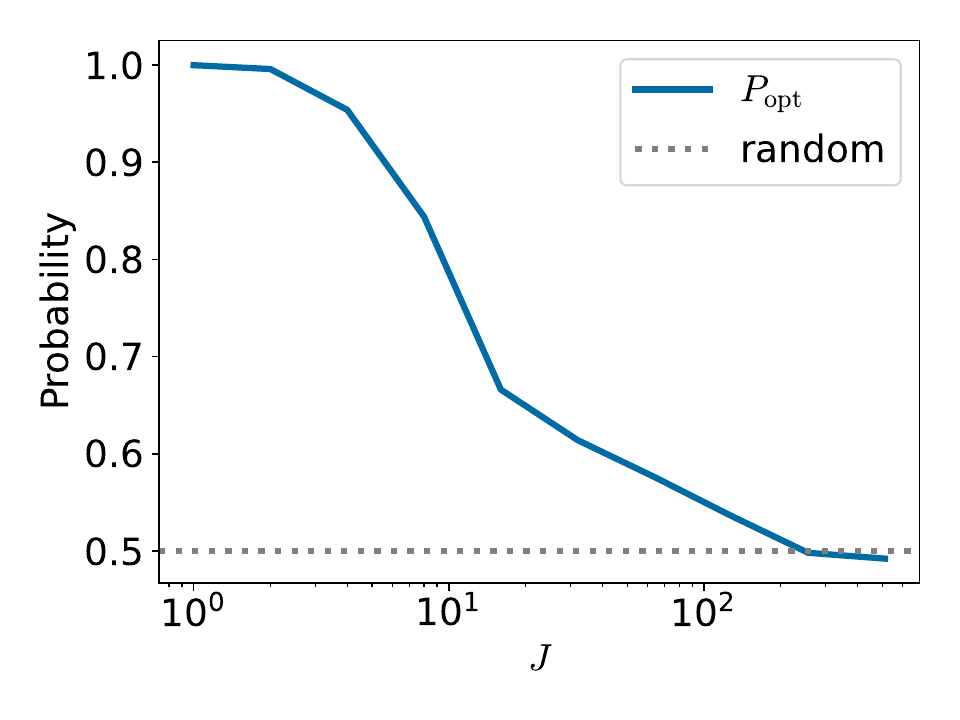}
    \caption{Probability $P_\mathrm{opt}$ to obtain ground states for the trivial problem \hl{versus} 
    coupling strength. 
    Baseline probability $0.5$ obtained by uniform sampling is labeled as `random'.}
    \label{fig:simple_ising}
\end{figure}

\subsection{Interaction-Extension Method}\label{subsec:experiment_iem}

\hl{We validate the IEM 
under minor-embedding through two experiments.
The first experiment is conducted on a trivial Ising problem, which serves to demonstrate the detrimental impact of a large dynamic range and the mechanism of the method on a minimal scale.
The second utilizes benchmark QUBO instances from MQLIB to ensure that the effect remains valid in a more realistic setting.
Specifically, we adopt instances where the external fields tend to be large to examine whether the coupling reduction remains the dominant factor for improving sample quality even in such challenging cases.}

\subsubsection{Trivial Problem}

First, we demonstrate how a large dynamic range harms the \hl{sample quality} of quantum annealers by considering the following trivial problem.
The Ising Hamiltonian is given as 
\begin{align}
    H = \sigma_1 \sigma_2 + \frac{1}{J}\sigma_2 \sigma_3
\end{align}
with a positive constant $J>0$.
The ground states are clearly given as $(\sigma_1, \sigma_2, \sigma_3) = (+1,-1,+1),(-1,+1,-1)$, regardless of the value of $J$.
It requires high numerical precision to accurately compute the ground state for large $J$.
If $J$ is too large, the quantum annealer cannot discriminate $H$ from the Hamiltonian \hl{$H_\infty = \sigma_1 \sigma_2$}, 
which has two more ground states, and returns wrong samples with 50\% probability.
\hl{To evaluate the precision of the quantum annealer,} we take 500 samples from the quantum annealer, 
varying $J$ from $1$ to $512$.
The probability $P_\mathrm{opt}$ of obtaining a correct ground state is shown in Fig.~\ref{fig:simple_ising}.
For $J\ge 256$, we get $P_\mathrm{opt}\sim 0.5$, which implies that the noise is dominant over the $1/J$ term.
Note that minor-embedding is not applied here since  $H$ can be directly embedded into the quantum annealer.

\hl{To verify the IEM,} we apply the method to the above Hamiltonian $H$ with $J=512$.
To ease the notation, we define a rescaled Hamiltonian:
\begin{align}
    H_\mathrm{rescaled} = J \sigma_1 \sigma_2 + \sigma_2 \sigma_3.
\end{align} 
We reduce the maximum coefficient from $J$ to $\hat J$ using the IEM for $\hat J = 32, 16, 8,4,2,1$.
For each resulting Hamiltonian, we run minor-embedding search with 10 different random seeds. 
We sample 100 solutions from the quantum annealer 
for each embedding, varying the chain strength from $\hat{J}/8$ to $128\hat{J}$.
The obtained samples are \hl{projected onto} 
the original Hamiltonian by \hl{discarding} the auxiliary variables.

\begin{figure}[t]
    \centering
    \includegraphics[width=0.94\linewidth]{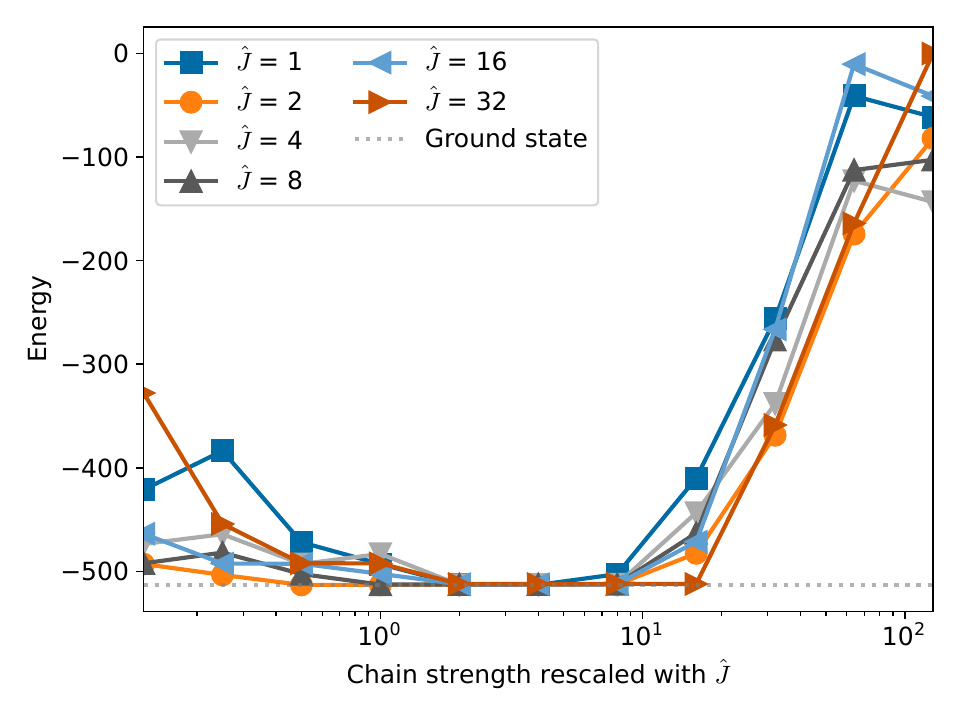}
    \caption{Average energy of samples on coefficient-reduced trivial problem.}
    \label{fig:simple_reduced_ising_energy}
\end{figure}

\begin{figure}[t]
    \centering
    \includegraphics[width=0.94\linewidth]{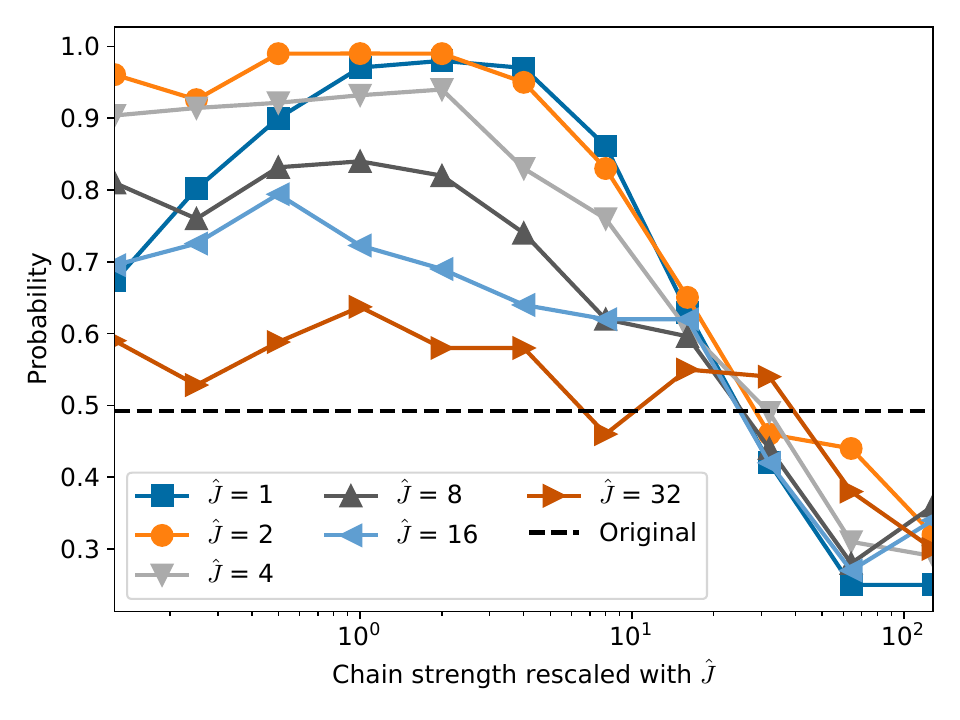}
    \caption{Probability of obtaining ground states on coefficient-reduced trivial problem. `Original' represents average energy of samples for non-reduced Hamiltonian $H$.}
    \label{fig:simple_reduced_ising_p_opt}
\end{figure}

Fig.~\ref{fig:simple_reduced_ising_energy} shows the average energy of samples \hl{against the chain strength}.
From the figure, the \hl{optimal} chain strength lies around $3 \hat J$ for all $\hat J$, suggesting that the chain strength can be reduced at the same rate as the coefficient in the logical Hamiltonian.
To observe the overall \hl{sample quality}, the probability $P_\mathrm{opt}$ of obtaining a ground state is shown in Fig.~\ref{fig:simple_reduced_ising_p_opt}.
We see that $P_\mathrm{opt}$ drastically \hl{increases} by reducing the maximum coefficient in $H_\mathrm{rescaled}$, reaching almost $P_\mathrm{opt}=1$ with $\hat{J} = 1$ or $2$.
The $\hat{J} = 1$ case yields a slightly \hl{lower value of $P_\mathrm{opt}$} than the $\hat{J} = 2$ case, possibly due to the substantial increase in the number of variables.
Note that the number of auxiliary variables for coefficient-reduction is given as $J/\hat{J} - 1$, which amounts to $511$ in the $\hat{J}=1$ case.

\subsubsection{MQLIB Instances}

We evaluate the IEM on the gka.*b instances, which are shown to have relatively large physical external field coefficients in Section~\ref{subsec:h_reduction_by_minor-embedding}, for * in $\{1,2,3\}$.
The numbers of variables in the instances are 20, 30, and 40, respectively.
The coupling coefficients in each problem were randomly sampled from an interval $[1, 50]$.
\hl{See the original paper~\cite{glover1998adaptive} for more detail on the instances}.
We reduce the maximum coefficient to $\hat J$ for $\hat J = 50, 40, 30, 20, 10$, where $\hat J = 50$ means no reduction is applied.
We run minor-embedding search with 10 different random seeds \hlx{for each $\hat J$ and sample 100 solutions for each embedding and chain strength.} 
The obtained samples are interpreted as samples for the original Hamiltonian by ignoring the auxiliary variables.

\hl{To investigate the reduction effect of the IEM on physical coupling coefficients and its impact on sample quality, we identify the optimal chain strength and evaluate the performance at that strength.
This experiment complements the scaling factor analysis 
in Section~\ref{subsec:h_reduction_by_minor-embedding}.
The optimal chain strength is estimated via a grid search, where the average energy of the samples is evaluated across various chain strengths.}
Note that the QUBO instances have a trivial solution $(0, \ldots, 0)$ with \hl{zero energy, whereas the optimal objective value is negative.
Since samples often yield large positive energies that far exceed the trivial solution, they can act as outliers that obscure the performance metric.
To mitigate this, we truncate positive energy values to zero before averaging.}

\begin{figure*}[t]
     \centering
     \subfloat[gka.1b\label{fig:pp_energy_gka1b}]{
         \includegraphics[width=0.32\textwidth]{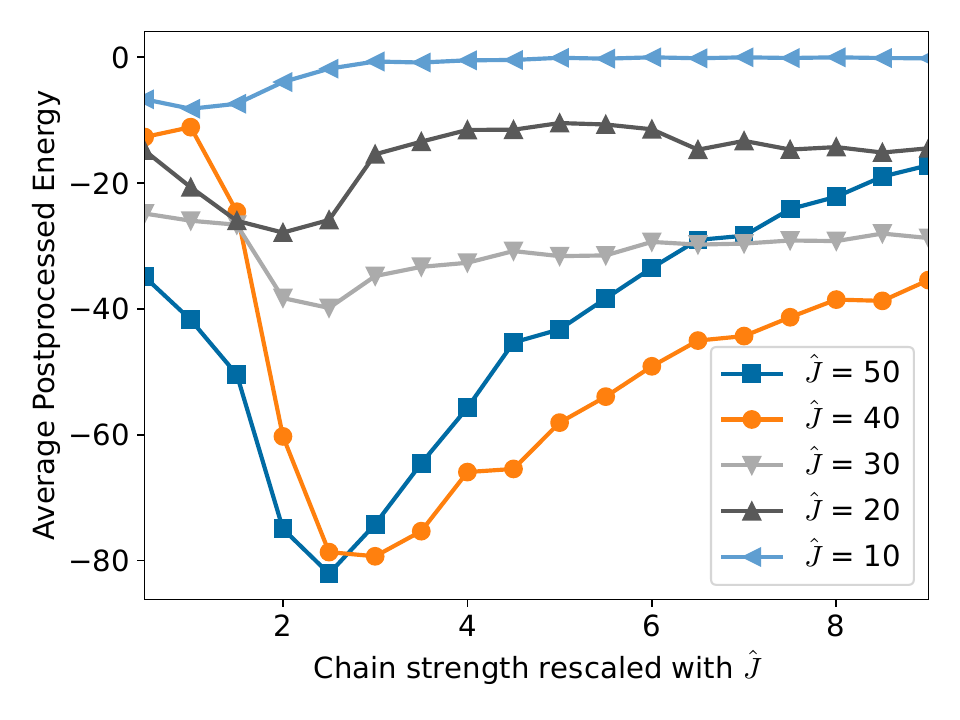}
     }
     \hfil
     \subfloat[gka.2b\label{fig:pp_energy_gka2b}]{
         \includegraphics[width=0.32\textwidth]{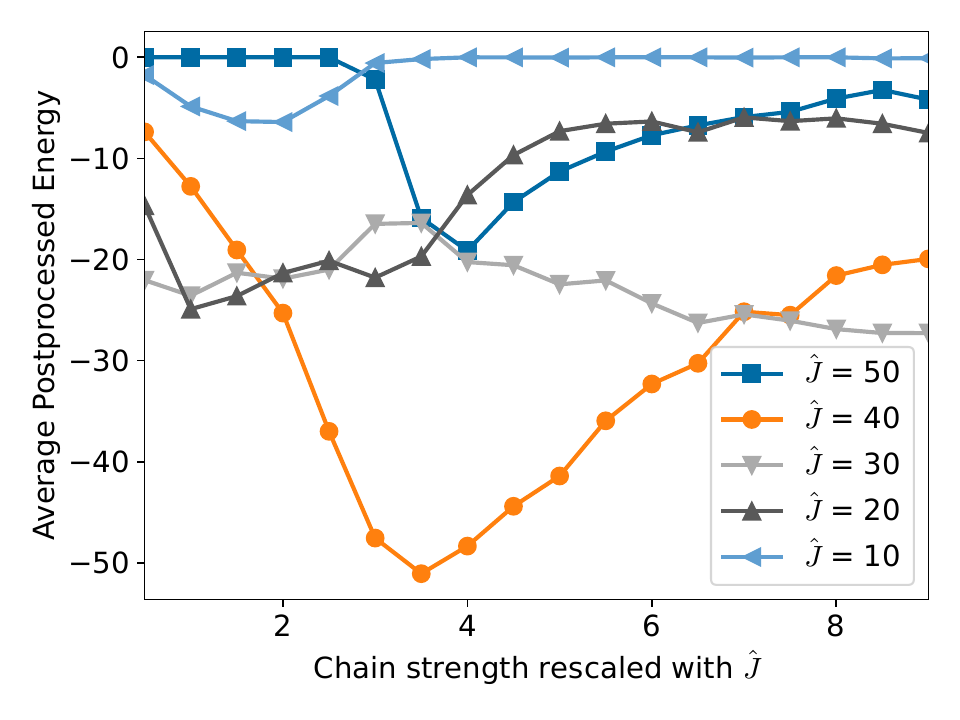}
     }
     \hfil
     \subfloat[gka.3b\label{fig:pp_energy_gka3b}]{
         \includegraphics[width=0.32\textwidth]{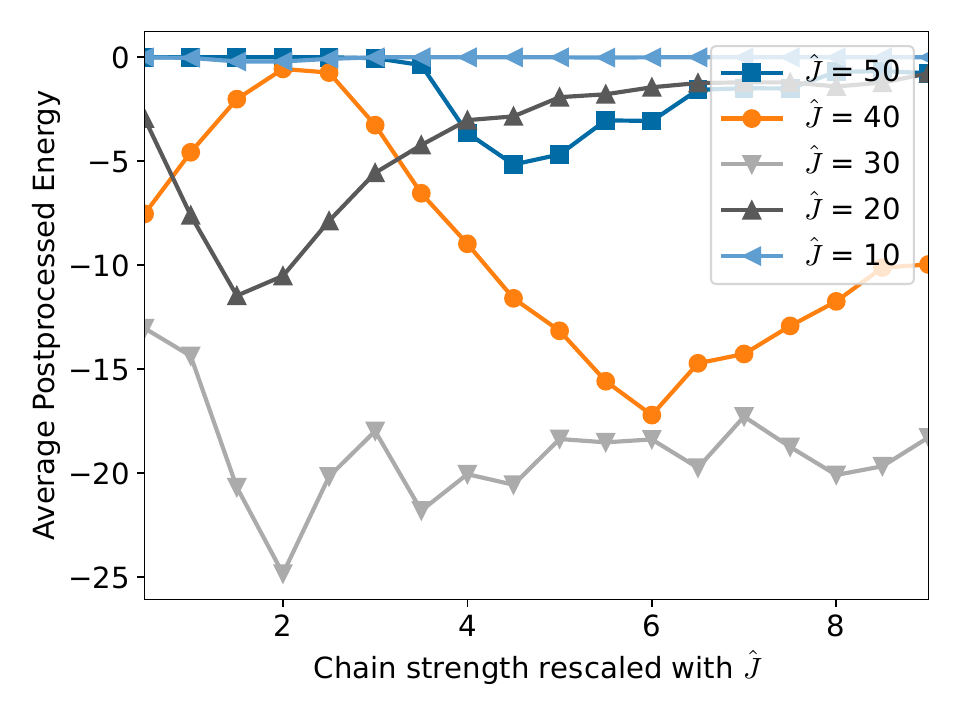}
     }
    \caption{
    Average energy on benchmark QUBO instances.
    }
    \label{fig:pp_energy_gka_b}
\end{figure*}

\hl{The average energy against the chain strength (rescaled by $\hat J$ for visibility) is shown in Fig.~\ref{fig:pp_energy_gka_b}.}
We observe that $\hat J=50, 40, 30$ result in the best average energy on gka.1b, gka.2b, and gka.3b, respectively, \hl{which implies that the IEM improves the sample quality} \hlx{on the latter two instances}.
Smaller $\hat J$ does not necessarily \hl{result in lower energy}, possibly 
due to the \hl{substantial} increase \hl{in} auxiliary variables.
\hl{Given that the original coefficients are uniformly distributed up to 50, approximately $1 - \hat{J}/50$ of the interactions exceed $\hat{J}$ and thus require auxiliary spins for each, leading to a significant variable overhead for small $\hat{J}$.}

\hl{From Fig.~\ref{fig:pp_energy_gka_b}, the optimal chain strength for the original coefficients $\hat J = 50$ is derived as $2.5 \hat J, 4\hat J$, and $4.5\hat J$ for gka.1b, gka.2b, and gka.3b, respectively, which yields the $s_{\hat J}/s_J$ column in Table~\ref{tab:gka_b_dr} of Exp.~1.}
On the smallest instance, gka.1b, the \hl{optimal} chain strength lies around $2.5 \hat J$ for $\hat J \ge 30$ and gets slightly smaller for $\hat J = 10, 20$.
This result suggests that the reduction rate of the \hl{optimal} chain strength is as high as that of the logical couplings.
The cases on gka.2b and gka.3b are more subtle, as the behavior of average energy is noisier.
By choosing $\hat J$ which gives the lowest average energy, i.e. $\hat J = 40$ on gka.2b and $\hat J = 30$ on gka.3b, we observe that the lowest energy is attained at chain strength $3.5 \hat J$ and $2 \hat J$, respectively,
which are smaller than that of the $\hat J = 50$ case, i.e., $4 \hat J$ and $4.5 \hat J$, respectively.
Overall, the coefficient reduction of logical couplings successfully reduces the \hl{optimal} chain strength 
at least for best-performing $\hat J$.

\begin{table}[t]
    \centering
    \caption{Scaling factors on gka.*b instances with coupling coefficient reduction.} 
    \label{tab:gka_b_dr_with_j_reduction}
    \begin{tabular}{c|r|r|r|r}
    Instance & size & $\hat J$ & $s_{\tilde h} / s_J$ & $s_{\tilde J}/ s_{J}$  \\
    \hline
    gka.1b   & 20   & 40 & $1.25$   & $1.50 $    \\
    gka.2b   & 30   & 40 & $1.18$   & $1.75 $       \\
    gka.3b   & 40   & 30 & $0.97$   & $1.00 $    
    \end{tabular}
\end{table}

\hl{In conclusion, the IEM is shown to be effective in improving sample quality by reducing the physical coupling coefficients across the test QUBO instances.
The data on the optimal chain strength for $\hat{J} = 50$ were obtained as a byproduct, which complements the findings in Section~\ref{subsec:h_reduction_by_minor-embedding}.}
Additionally, the scaling factors $s_{\tilde{h}}$ and $s_{\tilde{J}}$ of the physical external field and coupling \hl{for the best-performing $\hat{J}$ from the set $\{10, 20, 30, 40\}$, 
are summarized} in Table~\ref{tab:gka_b_dr_with_j_reduction}.
These results \hl{confirm} that the inequality $s_{\tilde{h}} \le s_{\tilde{J}}$ holds even for the coefficient-reduced Hamiltonian, \hl{supporting the claims} in Section~V-B \hl{that reducing logical external field coefficients is unnecessary also under coupling coefficient reduction}.

\subsection{Bounded-coefficient Integer Encoding}\label{subsec:experiment_bce}

\hl{We verify the effectiveness of the BCE 
under minor-embedding.
As with Exp.~2 in Section~\ref{subsec:experiment_iem}, two experiments are conducted: one on a trivial integer problem and the other on the benchmark MKP instances.}

\subsubsection{Trivial Problem}

We consider a trivial problem of minimizing $(z-1)^2$ under $0\le z \le 191, z \in \mathbb Z$ to demonstrate the effect of the BCE method on a minimal example.
The minimum objective value is clearly $0$.
The integer variable $z$ is represented by multiple spin variables using the BCE with $\mu \in \{ 64,32,16,8,4,2 \}$.
The $\mu = 64$ case corresponds to the usual binary \hlx{encoding}.
We run minor-embedding search for each $\mu$ and sample 1,000 solutions for each chain strength.

\begin{figure}[t]
    \centering
    \includegraphics[width=0.825\linewidth]{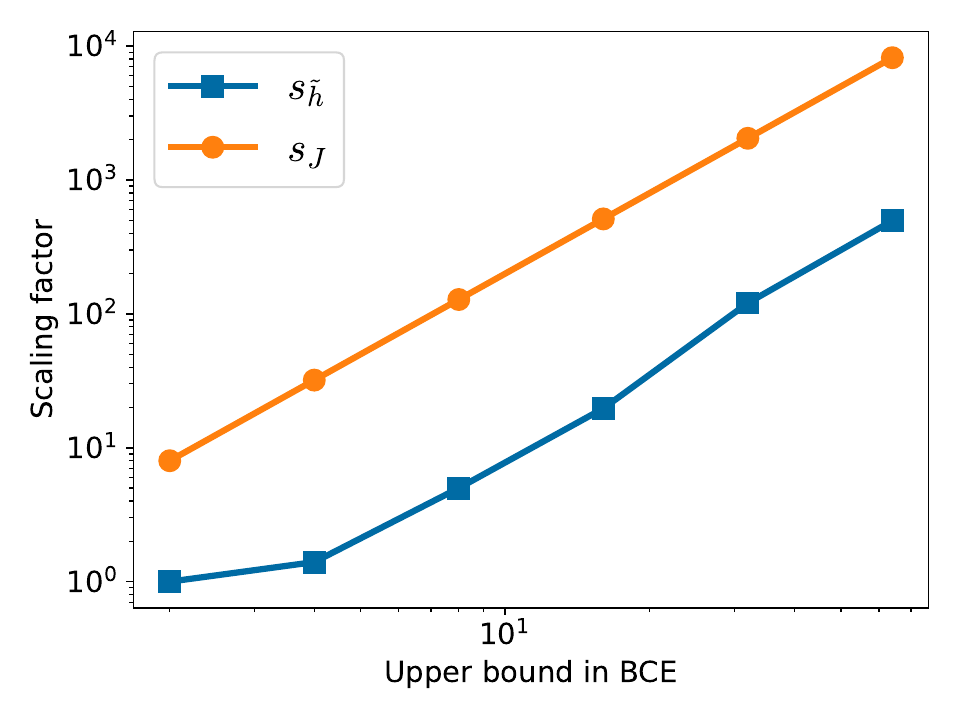}
    \caption{Scaling factors on trivial integer problem.}
    \label{fig:dr_trivial_integer_problem}
\end{figure}

We first observe the scaling factor $s_{\tilde{h}}$ and $s_{J}$ of the physical external field and logical coupling
to ensure that 
\hlx{the coupling strength is the bottleneck for the hardware precision}.
Fig.~\ref{fig:dr_trivial_integer_problem} shows the scaling factors for various $\mu$.
Note that $s_{J}$ scales quadratically with respect to $\mu$ by construction.
From the figure, we confirm that $s_{\tilde{h}} \le s_J$ holds for all $\mu$, which 
\hl{supports the claim in Section~\ref{subsec:h_reduction_by_minor-embedding} that the external fields do not require reduction on this problem.
Subsequently, the following evaluation focuses on the analysis of $s_{\tilde{J}}$ (i.e., the chain strength) and the resulting sample quality to verify the effectiveness of the method.}

\begin{figure}[t]
    \centering
    \includegraphics[width=0.82\linewidth]{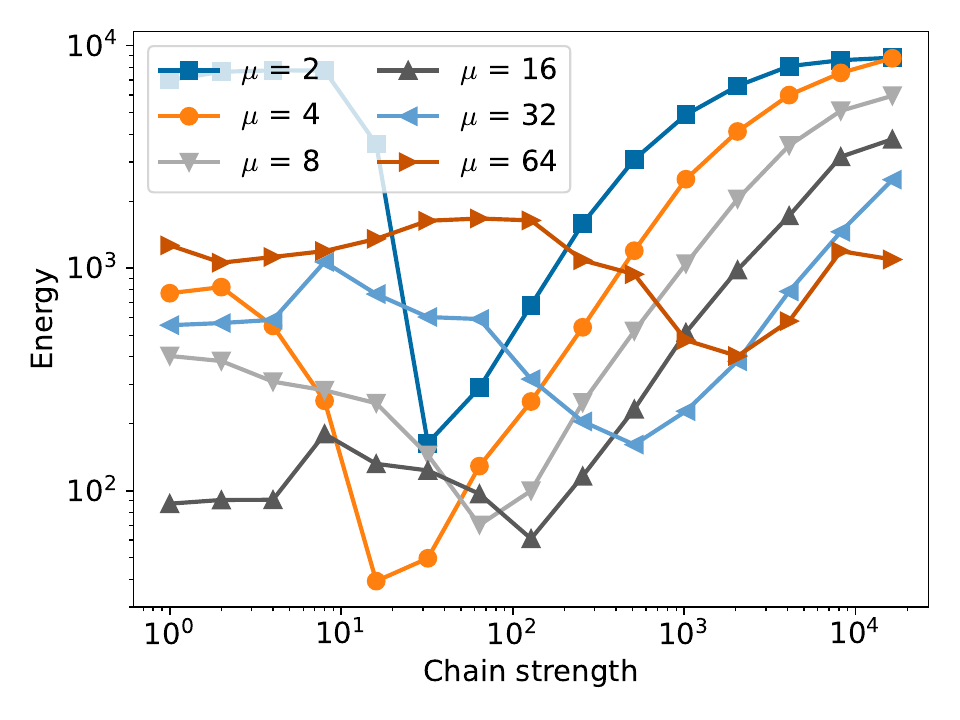}
    \caption{Average energy on trivial integer problem.}
    \label{fig:energy_trivial_integer_problem}
\end{figure}

\begin{figure}[t]
    \centering
    \includegraphics[width=0.825\linewidth]{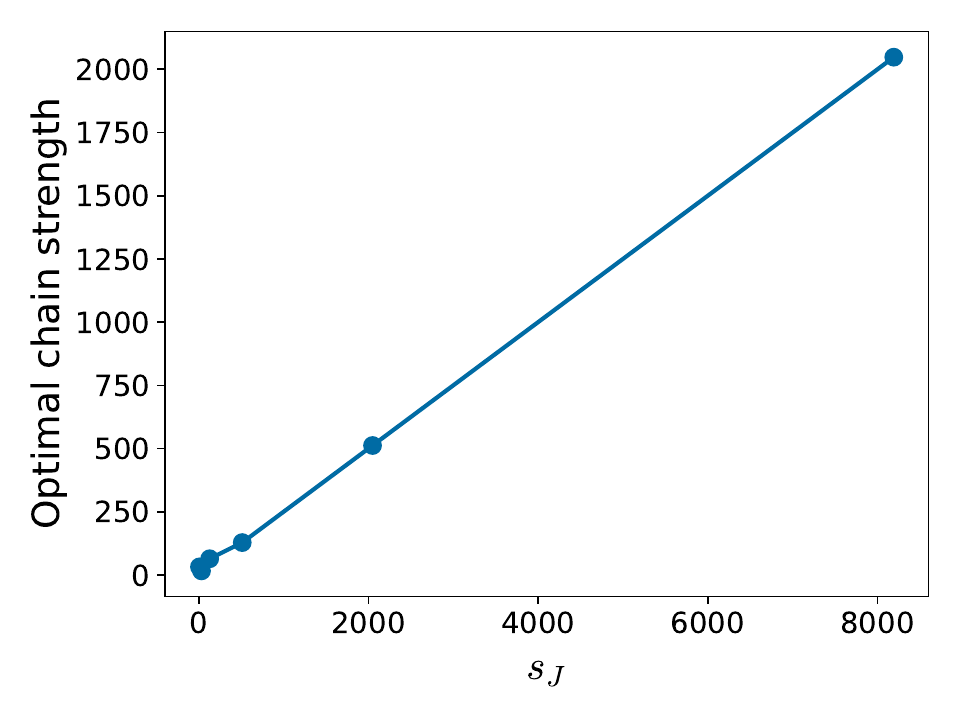}
    \caption{Relation between \hl{optimal} chain strength and scaling factor of logical coupling on trivial integer problem.}
    \label{fig:best_chain_trivial_integer_problem}
\end{figure}

\begin{figure}[t]
    \centering
    \includegraphics[width=0.825\linewidth]{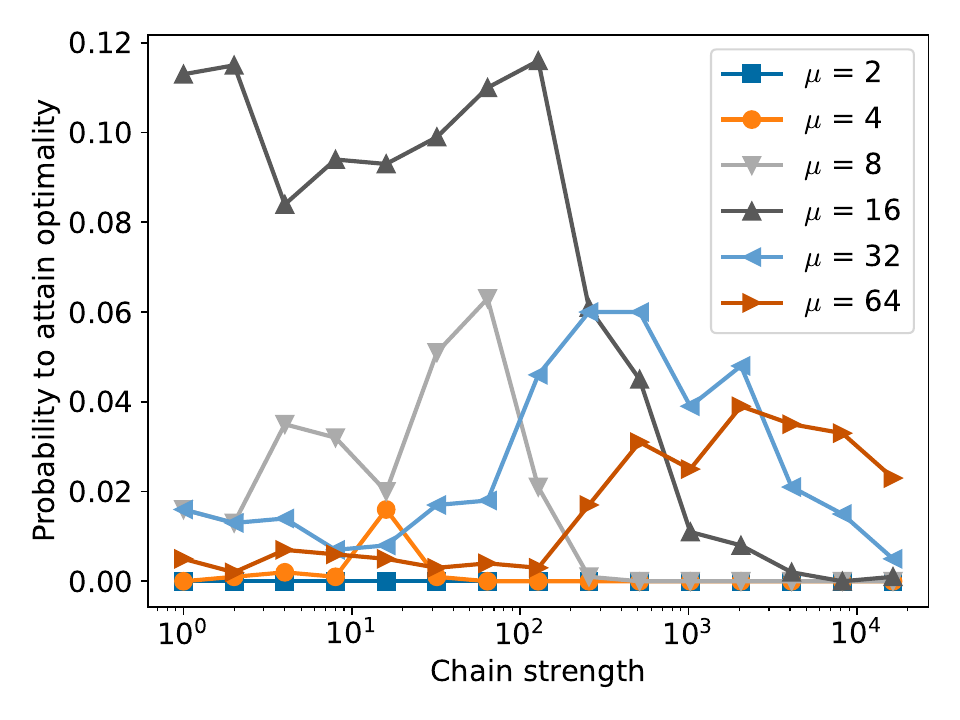}
    \caption{Probability of obtaining ground states on trivial integer problem.}
    \label{fig:p_opt_trivial_integer_problem}
\end{figure}

Next, we evaluate the reduction effect on the \hl{optimal} chain strength \hl{to observe the impact of the method on the physical coupling coefficients}.
Fig.~\ref{fig:energy_trivial_integer_problem} shows the average energy of samples \hl{against the chain strength}.
We observe a clear tendency that the \hl{optimal} chain strength is reduced as the coefficient bound $\mu$ decreases.
\hl{To observe the scaling,} we plot the \hl{optimal} chain strength attaining minimum average energy against the maximum coupling coefficient $s_J$ in the logical Hamiltonian in Fig.~\ref{fig:best_chain_trivial_integer_problem}.
The linear relation between them illustrates that BCE effectively reduces the chain strength as well as the logical couplings on this trivial problem.

To evaluate the overall \hl{sample quality}, the probability $P_\mathrm{opt}$ of obtaining the ground state is presented in Fig.~\ref{fig:p_opt_trivial_integer_problem}.
It shows that $\mu \in [8, 32]$ successfully improves the \hl{sample quality} when the chain strength is suitably chosen.
In particular, $\mu=16$ significantly increases $P_\mathrm{opt}$ to more than 10\%.
On the other hand, the smallest upper bound $\mu=2$ does not produce the best average energy nor $P_\mathrm{opt}$.
We consider this is because setting $\mu$ to such a small value significantly increases the number of variables, which degrades the quality of minor-embedding and quantum annealing.
\hl{Note that the number of \hlx{logical} variables is $\lfloor 191 / \mu \rfloor + \log_2 \mu$, following Eq.~(\ref{eq:bce_notation}).}
Interestingly, the $\mu = 4$ case obtains the best average energy while leading to lower $P_\mathrm{opt}$ than even $\mu = 64$.
This is probably because the average accuracy on each qubit improves due to the coefficient reduction, while the increase in the number of variables exponentially decreases the probability of all variables taking their correct values simultaneously.


In summary, on the trivial integer problem, \hl{the BCE improves the sample quality by reducing the chain strength in an ideal way.
Moreover, the external field coefficients are sufficiently reduced by minor-embedding.}

\subsubsection{Multi-dimensional Knapsack Problem}

To assess the effect of the method on practical problems, we take the multi-dimensional knapsack problem (MKP) as an example.
The MKP is defined as follows:
\begin{align}\label{eq:mkp}
    \mathrm{Maximize\ }& \sum_{j=1}^n p_j \hlmx{x_j} \\
    \mathrm{subject\ to\ }& \sum_{j=1}^n w_{ij} x_j \le C_i \quad \mathrm{for\ } i=1,\ldots,m \\
    & x_j \in \{0,1\} \quad \mathrm{for\ } j=1,\ldots,n
\end{align}
where $n, m, p_j$, and $C_i$ are positive integers and $w_{ij}$ are non-negative integers.
Penalty terms representing the constraints are constructed by introducing $m$ integer variables $z_1, \ldots, z_m$ satisfying $0 \le z_i \le C_i$.
By flipping the sign of the objective, we obtain an unconstrained problem
\begin{align}\label{eq:mkp_penalized}
    \mathrm{Minimize\ }& -\sum_{j=1}^n p_j \hlmx{x_j} + \sum_{i=1}^m \lambda \left( \sum_{j=1}^n w_{ij} x_j - z_i \right)^2 \\
    \mathrm{subject\ to\ }& z_i \in \{0,1,\ldots, C_i\} \quad \mathrm{for\ } i=1,\ldots,m \\
    & x_j \in \{0,1\} \quad \mathrm{for\ } j=1,\ldots,n
\end{align}
For a given integer value $\mu > 0$, we obtain a QUBO \hl{formulation} by expanding each $z_i$ \hl{with multiple binary variables $y_{il} \in \{0,1\}$ as $z_i = \sum_l a_{il} y_{il}$ following Eq.~(\ref{eq:bce_coefficient}) in Section~\ref{subsec:bce}, which satisfies $a_{il} \le \mu$ and $\sum_l a_{il} = C_i$.}
For simplicity, the penalty coefficient $\lambda$ and upper bound $\mu$ are set independently from $i$ in our experiment.

\begin{figure}[t]
     \centering
     \subfloat[weing1\label{fig:dr_mkp_weing1}]{
         \includegraphics[width=0.81\linewidth]{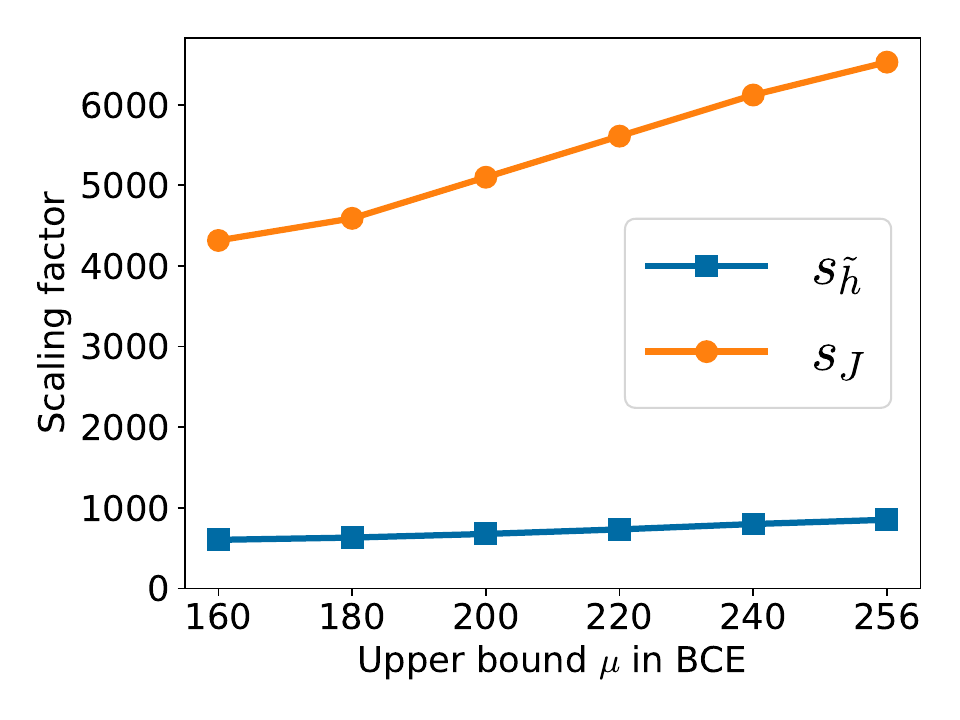}
     }
     \\ 
     \subfloat[weish06\label{fig:dr_mkp_weish06}]{
         \includegraphics[width=0.81\linewidth]{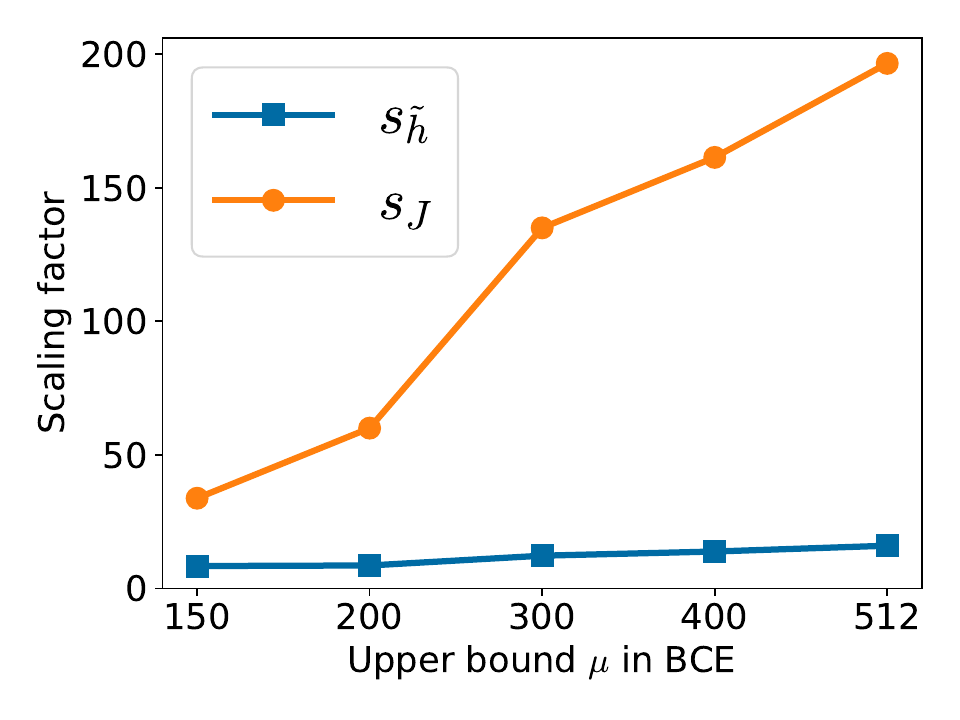}
     }
    \caption{
    Scaling factors on MKP instances.
    }
    \label{fig:dr_mkp}
\end{figure}

We take two MKP instances from OR-Library~\cite{beasley1990or}.
One is named 
weing1~\cite{weingartner1967methods}, which has 28 variables and 2 constraints.
The other is weish06~\cite{shih1979branch}, which has 40 variables and 5 constraints.
The penalty coefficient is set to $\lambda=0.3$ and $\lambda=0.003$ on weing1 and weish06, respectively.
The upper bound $\mu$ is varied from 160 to 256 on weing1 and from 100 to 512 on weish06.
On both instances, the maximum $\mu$ corresponds to the binary \hlx{encoding}. 
The values of $\lambda$ and ranges of $\mu$ are determined on the basis of preliminary experiments, see Appendix~\ref{app:mkp_preliminary} for details.
For each $\mu$, we run minor-embedding search for 10 different random seeds.
For each embedding and chain strength, we \hl{collect} 100 samples from the quantum annealer with annealing time 1~ms.

\hl{
To ensure that the hardware precision is constrained by the coupling strength,}
we observe the scaling factor $s_{\tilde{h}}$ and $s_{J}$ of the physical external field and logical coupling.
We average $s_{\tilde{h}}$ over 10 minor-embeddings.
Fig.~\ref{fig:dr_mkp} shows the scaling factors for various $\mu$.
From the figure, we confirm that $s_{\tilde{h}} \le s_J$ holds for all $\mu$, \hl{which indicates that the external fields do not limit the hardware precision for this problem. 
This observation also reinforces the claim in Section~\ref{subsec:h_reduction_by_minor-embedding} that logical external field reduction is unnecessary.}

\begin{figure}[t]
     \centering
     \subfloat[weing1\label{fig:energy_mkp_weing1}]{
         \includegraphics[width=0.81\linewidth]{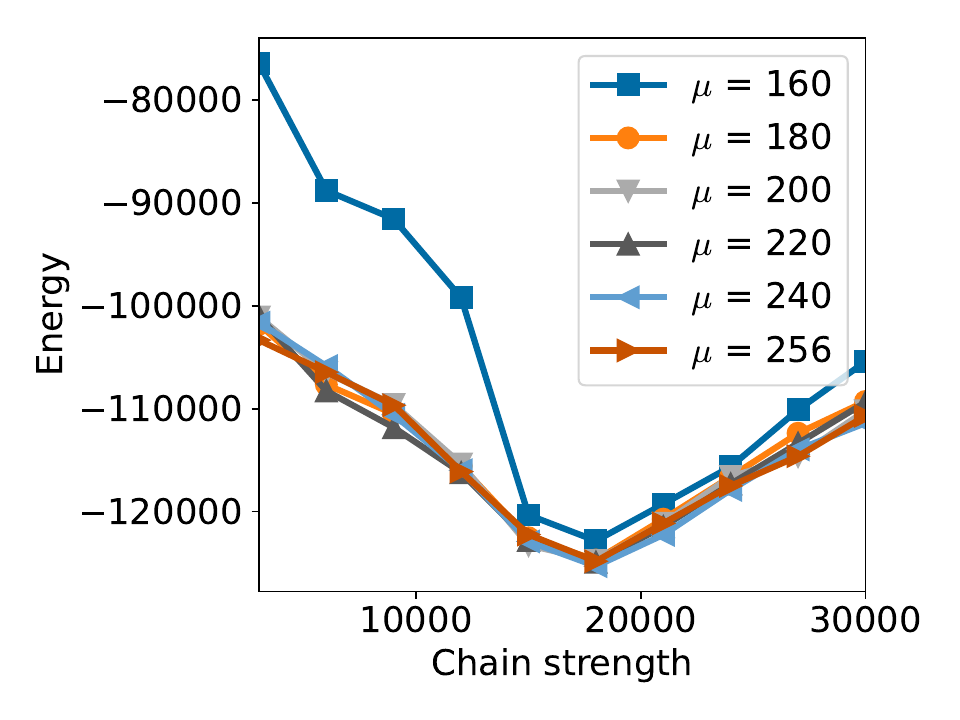}
     }
     \\ 
     \subfloat[weish06\label{fig:energy_mkp_weish06}]{
         \includegraphics[width=0.81\linewidth]{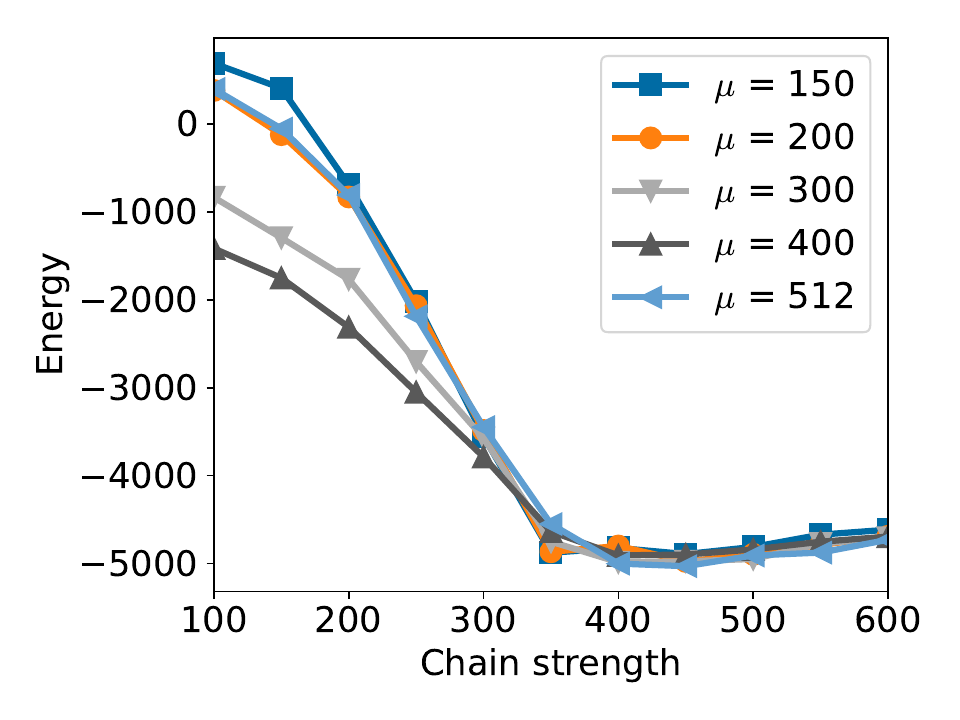}
     }
    \caption{
    Average energy on MKP instances.
    }
    \label{fig:energy_mkp}
\end{figure}

\begin{figure}[t]
     \centering
     \subfloat[weing1\label{fig:objective_mkp_weing1}]{
         \includegraphics[width=0.81\linewidth]{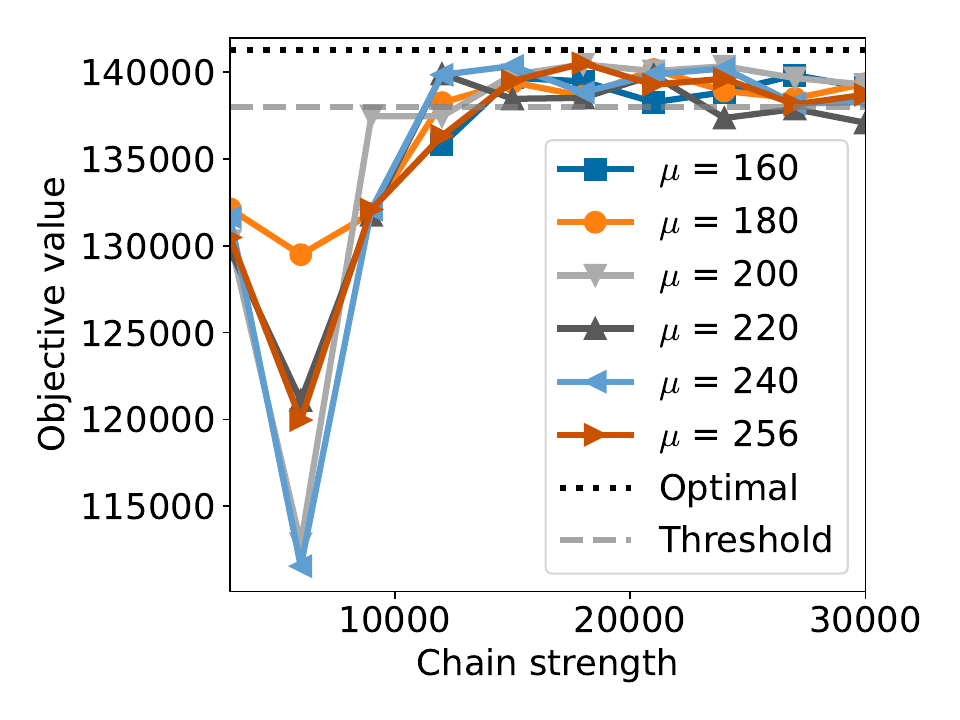}
     }
     \\ 
     \subfloat[weish06\label{fig:objective_mkp_weish06}]{
         \includegraphics[width=0.81\linewidth]{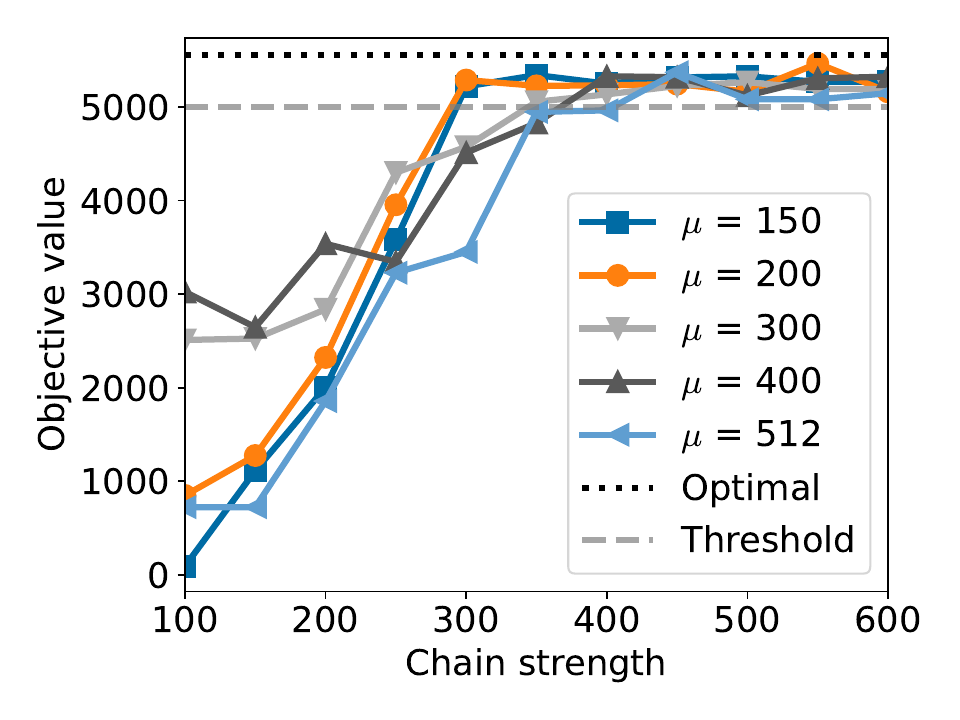}
     }
    \caption{
    Objective value on MKP instances.
    }
    \label{fig:objective_mkp}
\end{figure}

\begin{figure}[t]
     \centering
     \subfloat[weing1\label{fig:best_chain_mkp_weing1}]{
         \includegraphics[width=0.81\linewidth]{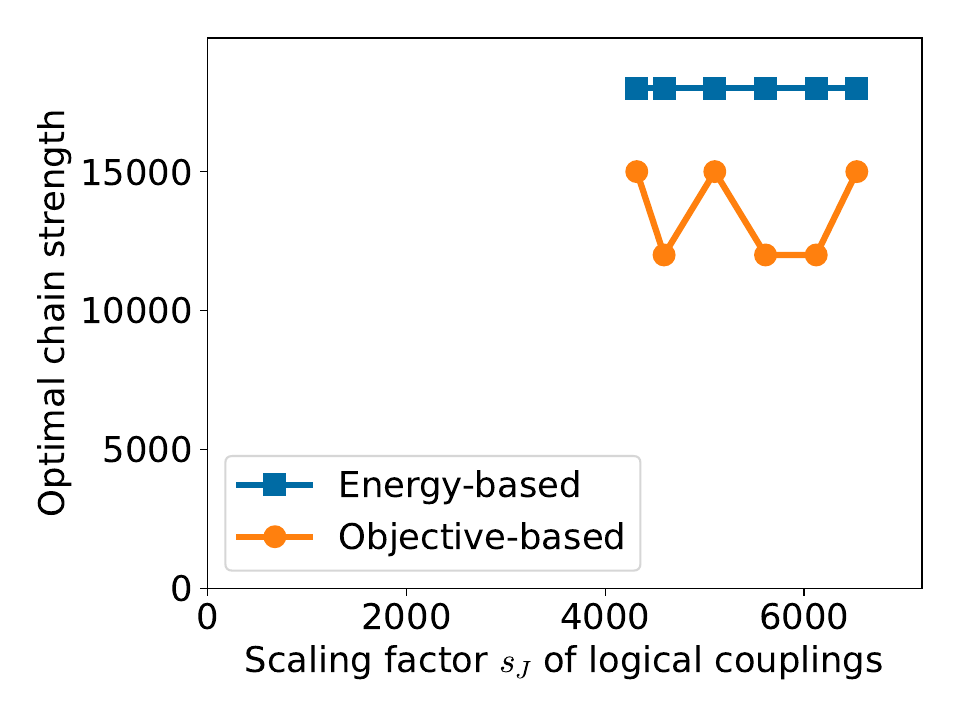}
     }
     \\ 
     \subfloat[weish06\label{fig:best_chain_mkp_weish06}]{
         \includegraphics[width=0.81\linewidth]{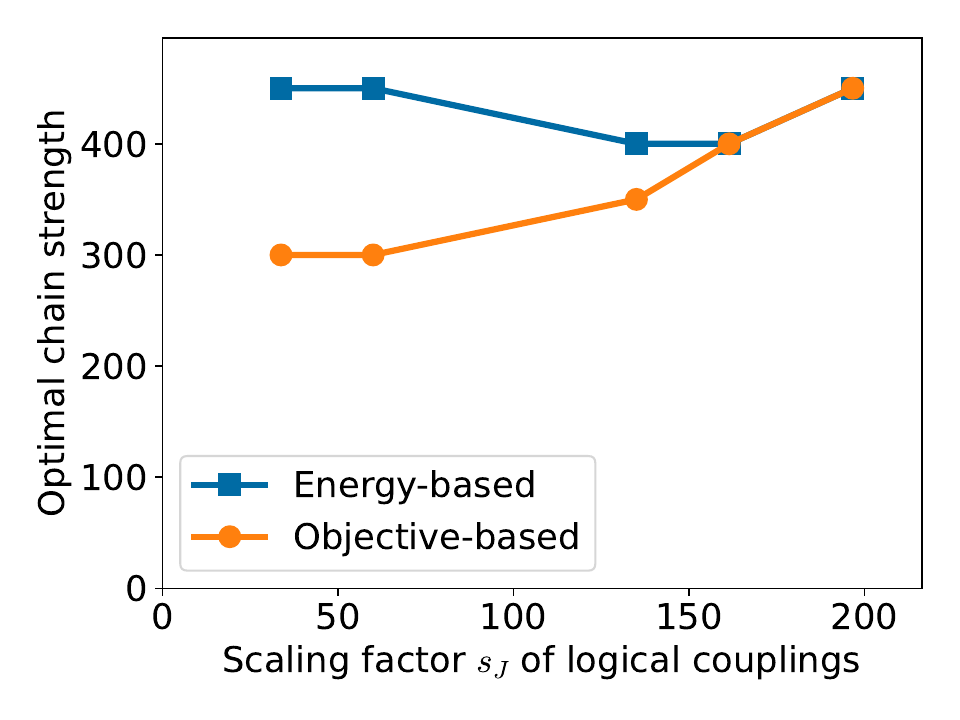}
     }
    \caption{
    \hl{Optimal} chain strength on MKP instances.
    }
    \label{fig:best_chain_mkp}
\end{figure}

\hl{Next, we evaluate the BCE using samples from the quantum annealer.}
On the MKP, 
the sample quality is evaluated by two metrics: the energy and the \hl{original} objective value, \hl{which correspond to the objective function in Eq.~(\ref{eq:mkp_penalized}) and Eq.~(\ref{eq:mkp}), respectively.}
The energy is suited to measure the bare performance of quantum annealers, while the objective value \hl{of feasible solutions} is a more practical metric.
We estimate the \hl{optimal} chain strength with \hlx{respective} metrics.

\hl{To assess the optimal chain strength and sample quality,}
the average energy of samples and the highest objective values over feasible solutions 
are shown in Fig.~\ref{fig:energy_mkp} and Fig.~\ref{fig:objective_mkp}, respectively.
\hlx{Note that a large chain strength is required for weing1 to suppress chain breaks due to the large logical coupling coefficients (Fig.~\ref{fig:dr_mkp}).}
We observe that the objective value reaches a plateau for sufficiently large chain strength. 
We define the \hl{optimal} chain strength based on the objective value as the smallest chain strength among those reaching the plateau.
Specifically, we pick the chain strength with which the objective value exceeds the following heuristic thresholds: 138,000 on weing1 and 5,000 on weish06.
From Fig.~\ref{fig:best_chain_mkp}, 
the \hl{optimal} chain strength based on the energy does not change significantly on both instances.
The \hl{optimal} chain strength based on the objective value is not changed much either on weing1.
It is reduced by a certain factor on weish06, but the reduction rate is relatively small compared with that of the logical coupling coefficients as shown in Fig.~\ref{fig:best_chain_mkp_weish06}.
\hlx{Accordingly, the overall improvement in \hl{sample quality} is not observed from Fig.~\ref{fig:energy_mkp} and Fig.~\ref{fig:objective_mkp}}.

From the results, we conclude that the effect of the BCE is limited on the MKP, which is in contrast with the trivial integer problem.
Although the gap would be attributed to the existence of other quadratic terms containing $x_j$ \hlx{that appear} in the QUBO objective function Eq.~(\ref{eq:mkp_penalized}), the precise \hl{mechanism remains elusive and warrants further investigation.}

\subsection{Augmented Lagrangian Method}\label{subsec:experiment_augmented_lagrangian}

\hlbegin
We verify the impact of the ALM, or penalty perturbation, on the coefficient reduction for the physical Hamiltonian to enhance sample quality.
The quadratic assignment problem (QAP) is employed 
for this evaluation, as its quadratic objective function is inherently compatible with Ising models, and its mathematical structure allows for a methodologically transparent application of penalty perturbation.

Ideally, as with the chain strength, the reduction effect on the \hl{optimal} penalty coefficient should be quantitatively measured to validate the method.
However, identifying the \hl{optimal} penalty coefficient is not straightforward, as it involves the fundamental trade-off between the feasibility and objective value of samples.
Therefore, in our experiments, we first observe the qualitative effect of the method using simulated annealing (SA) in place of quantum annealing, to establish a baseline without minor-embedding.
We then evaluate the method by testing whether a consistent reduction effect is maintained when minor-embedding is applied.
\hlend

\begin{figure}[t]
     \centering
     \subfloat[nug5\label{fig:feasibility_qap_nug5_sa}]{
         \includegraphics[width=0.81\linewidth]{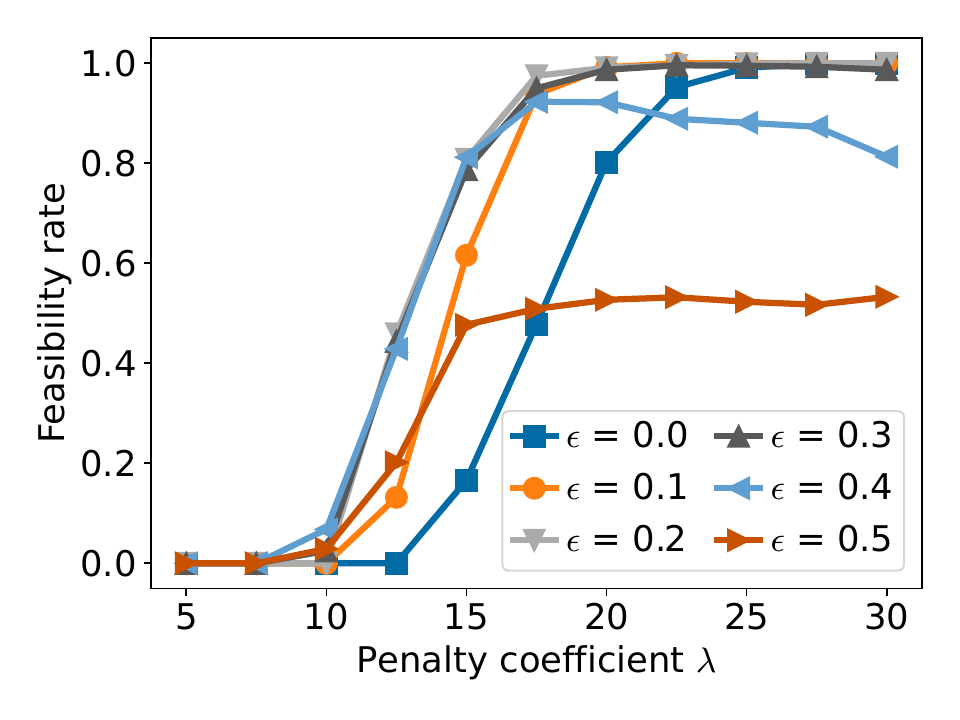}
     }
     \\ 
     \subfloat[tai5a\label{fig:feasibility_qap_tai5a_sa}]{
         \includegraphics[width=0.81\linewidth]{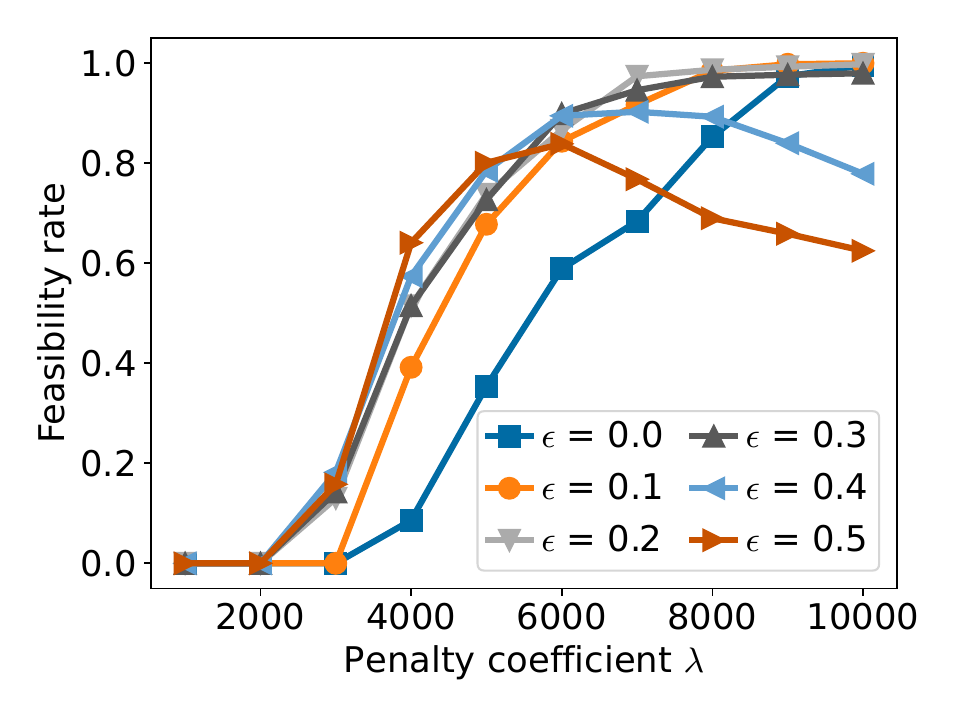}
     }
    \caption{
    Feasibility rate on QAP \hl{instances using} SA.
    }
    \label{fig:feasibility_qap_sa}
\end{figure}

\begin{figure}[t]
     \centering
     \subfloat[nug5\label{fig:objective_qap_nug5_sa}]{
         \includegraphics[width=0.81\linewidth]{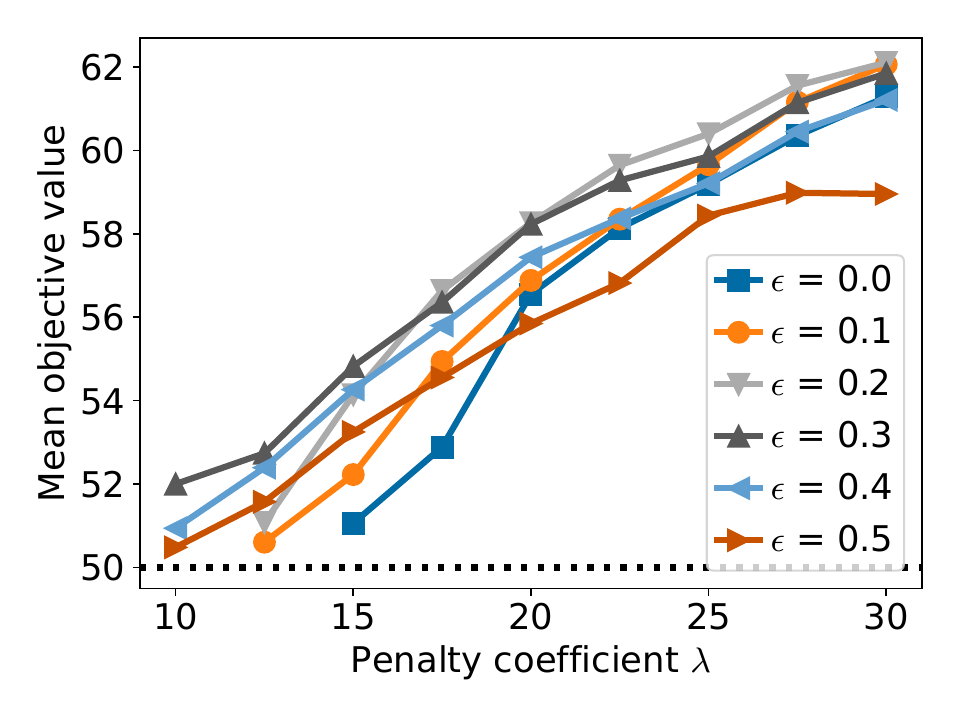}
     }
     \\ 
     \subfloat[tai5a\label{fig:objective_qap_tai5a_sa}]{
         \includegraphics[width=0.81\linewidth]{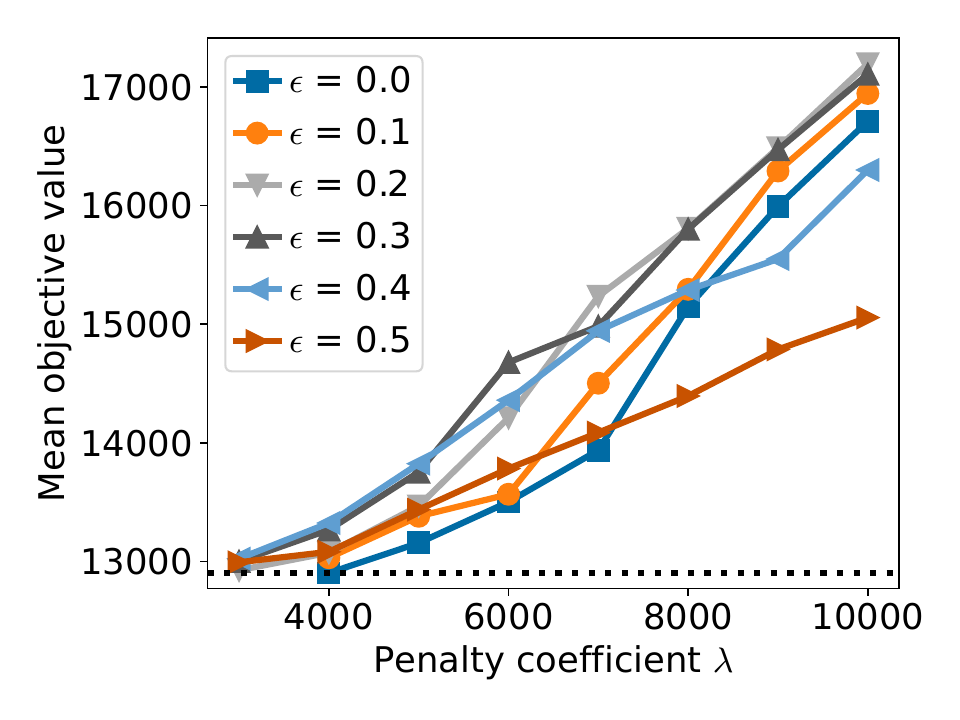}
     }
    \caption{
    Mean objective values on QAP \hl{instances using} SA. Dotted line represents optimum.
    }
    \label{fig:objective_qap_sa}
\end{figure}

\subsubsection{Problem Formulation and Penalty Perturbation}

\hl{For an integer $n$ and two sets of non-negative values $f_{ij}$ and $d_{ij}$ ($i,j=1,\ldots,n$), the QAP is defined as}
\begin{align}
    \mathrm{Minimize\ }& \sum_{i=1}^n\sum_{j=1}^n\sum_{k=1}^n\sum_{l=1}^n f_{ij}d_{kl} x_{ik}x_{jl} \\
    \mathrm{subject\ to\ }& \sum_{i=1}^n x_{ij} = 1,  \quad \mathrm{for\ } j=1,\ldots,n \label{eq:2way_1hot_1}\\
    & \sum_{j=1}^n x_{ij} = 1, \quad \mathrm{for\ } i = 1,\ldots,n \label{eq:2way_1hot_2}\\
    & x_{ij} \in \{0,1\} \quad \mathrm{for\ } i,j=1,\ldots,n.
\end{align}
The constraints Eq.~(\ref{eq:2way_1hot_1}) and Eq.~(\ref{eq:2way_1hot_2}) impose the matrix $(x_{ij})_{ij}$ to be a permutation matrix.
We use two small QAP instances\footnote{Available on http://mistic.heig-vd.ch/taillard/problemes.dir/qap.dir/qap.html.} called nug5~\cite{nugent1968experimental} and tai5a~\cite{taillard1991robust}.
The number of binary variables is $25$ on both instances.
The QAP can be translated into a QUBO form by converting the constraint conditions into penalty terms.
To assess the coefficient-reduction effect of the ALM, we introduce the following QUBO formulation with perturbation $\epsilon$ of constraints:
\begin{align}\label{eq:qap_penalized}
    \mathrm{Minimize\ }& \sum_{i=1}^n\sum_{j=1}^n\sum_{k=1}^n\sum_{l=1}^n f_{ij}d_{kl} x_{ik}x_{jl} \notag \\
    & + \sum_{j=1}^n \lambda \left( \sum_{i=1}^n x_{ij} - 1 - \epsilon \right)^2 \notag \\
    &+ \sum_{i=1}^n \lambda \left( \sum_{j=1}^n x_{ij} - 1 - \epsilon \right)^2 \notag \\
    &- 2 n \lambda \epsilon^2\\
    \mathrm{subject\ to\ }& x_{ij} \in \{0,1\} \quad \mathrm{for\ } i,j=1,\ldots,n.
\end{align}
Note that if we do not impose the penalty, i.e., $\lambda=0$, then the solution trivially becomes all-zero $x=(0,\ldots,0)$.
\hlx{Therefore,} as described in Section~\ref{subsec:augmented_lagrangian},
a positive perturbation $\epsilon \in (0, 0.5)$ effectively increases the penalty 
without increasing the penalty coefficient $\lambda$.
\hlx{The perturbation width $\epsilon$ is varied} for $\epsilon \in \{0, 0.1, 0.2, 0.3, 0.4, 0.5\}$ to observe the effect on the penalty coefficient $\lambda$ and the \hl{sample quality}.

\subsubsection{Validation without minor-embedding using SA}

\hl{To establish a baseline}, we demonstrate the effectiveness of the ALM in reducing penalty coefficients in a setting without minor-embedding.
\hl{SA for QUBO, implemented in D-Wave Ocean SDK, is used with default parameters to \hlx{collect} 1,000 samples for each $(\lambda,\epsilon)$ pair.}

\hl{First, the reduction effect is observed through feasibility rates.}
Fig.~\ref{fig:feasibility_qap_sa} illustrates the rate of feasible solutions \hl{as a function of the penalty coefficient}.
The results show that feasible solutions are obtained \hlx{with} smaller penalty coefficients when positive perturbation is applied.
Comparing \hlx{the results for $\epsilon=0.0$ and $\epsilon=0.3$, the penalty coefficient required to maintain a comparable feasibility rate decreases by approximately 30\%} on both instances.
When $\epsilon$ gets even larger, the maximum of the feasibility rate decreases.
\hl{This occurs because intense perturbation increases the likelihood of solutions violating the constraint with $\sum_i x_i \ge 2$.}

\hl{The effect is also evident in the objective values.}
Fig.~\ref{fig:objective_qap_sa} shows the average objective values over all feasible solutions.
In all cases, smaller penalty coefficients produce better objective values, \hl{provided that feasibility is maintained. 
Notably, with positive perturbation, the ALM achieves comparable or improved objective values at smaller penalty coefficients compared to the $\epsilon = 0$ case.}

\subsubsection{Validation with minor-embedding}


\begin{figure}[t]
     \centering
     \subfloat[nug5\label{fig:dr_qap_nug5}]{
         \includegraphics[width=0.81\linewidth]{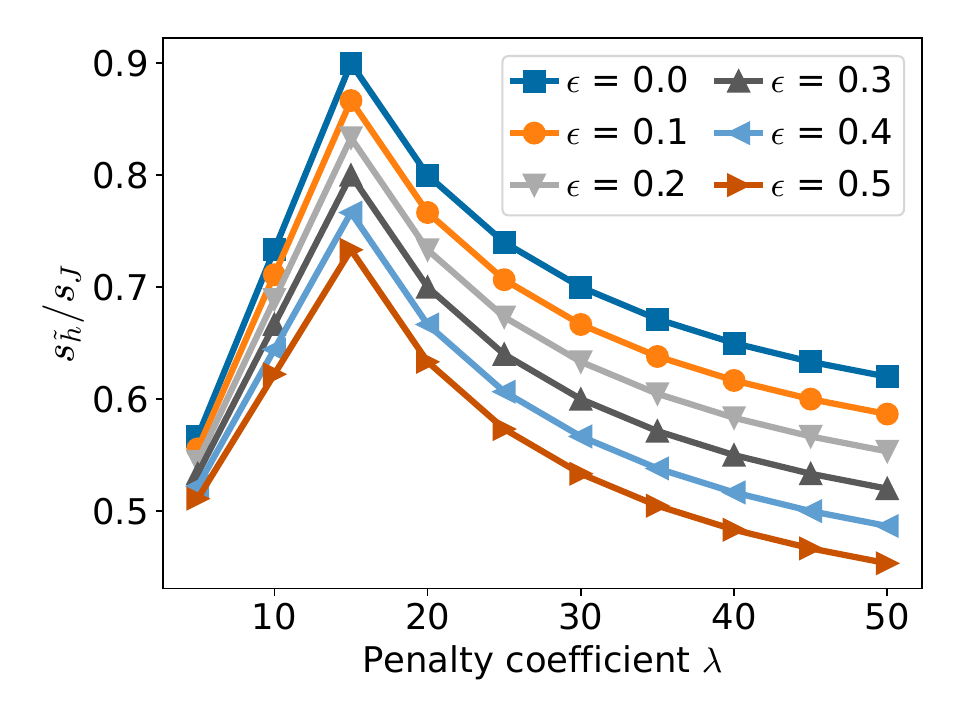}
     }
     \\ 
     \subfloat[tai5a\label{fig:dr_qap_tai5a}]{
         \includegraphics[width=0.81\linewidth]{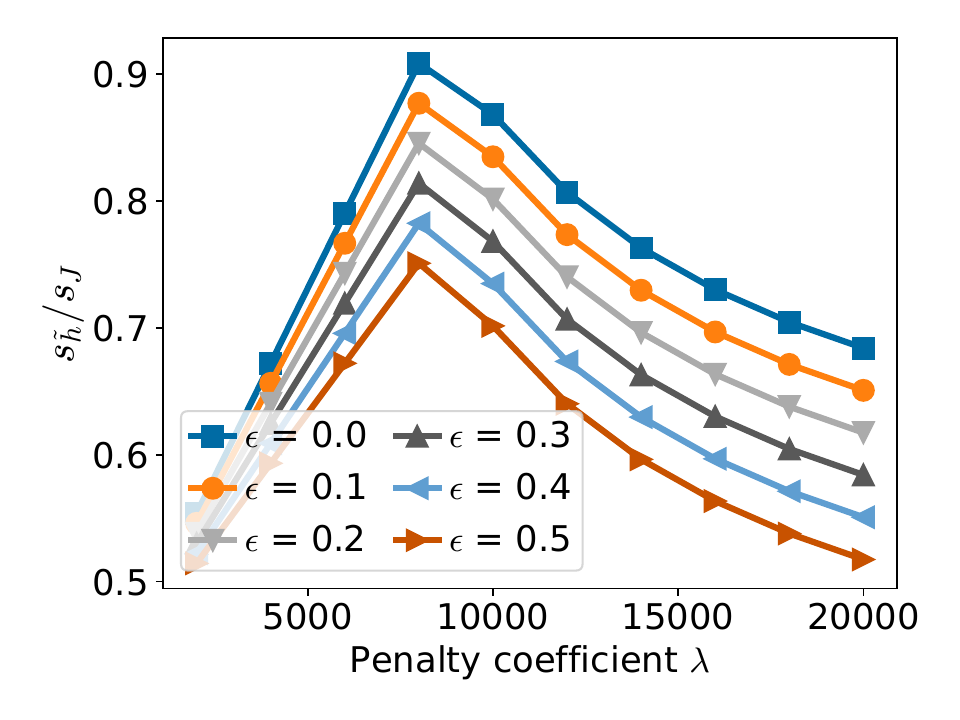}
     }
    \caption{
    Ratio of scaling factors on QAP \hl{instances}.
    }
    \label{fig:dr_qap}
\end{figure}

\begin{figure}[t]
     \centering
     \subfloat[nug5\label{fig:best_chain_qap_nug5}]{
         \includegraphics[width=0.81\linewidth]{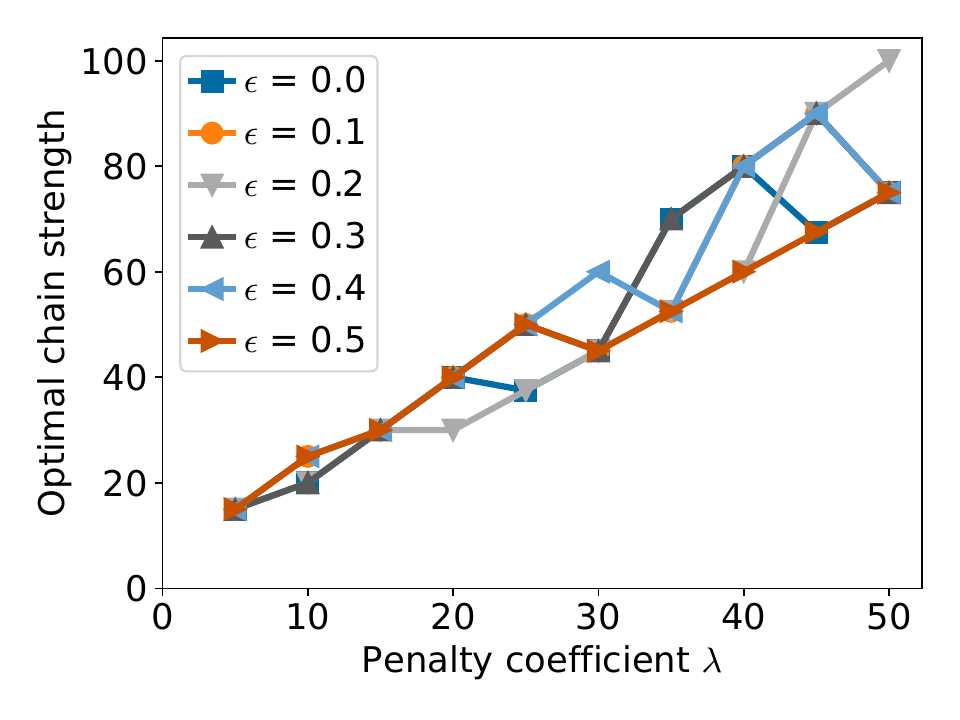}
     }
     \\ 
     \subfloat[tai5a\label{fig:best_chain_qap_tai5a}]{
         \includegraphics[width=0.81\linewidth]{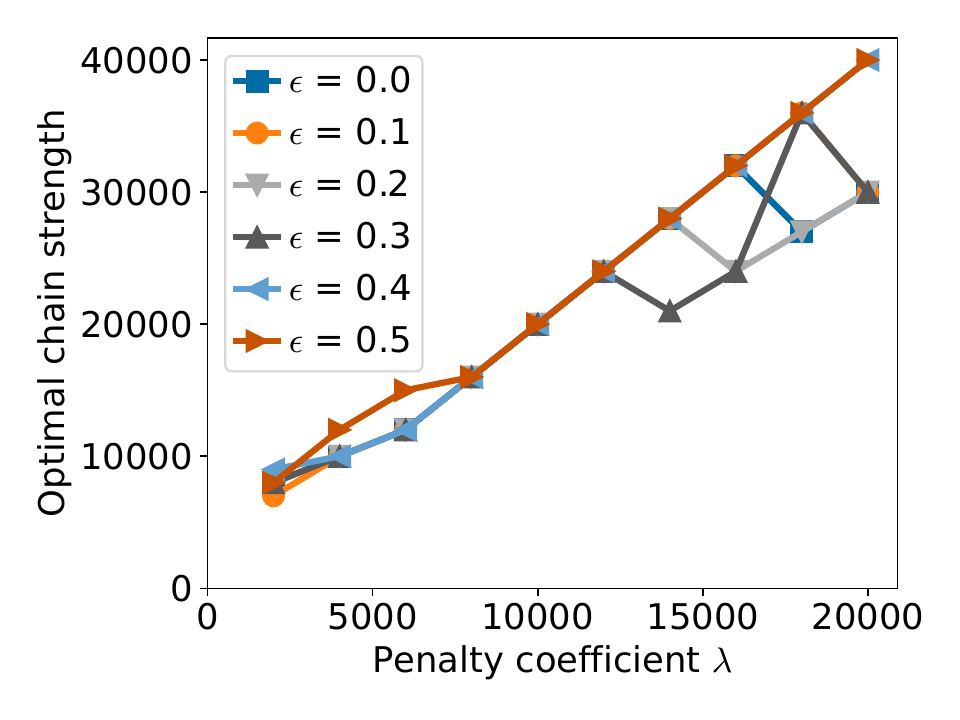}
     }
    \caption{
    \hl{Optimal} chain strength on QAP \hl{instances}.
    }
    \label{fig:best_chain_qap}
\end{figure}

\begin{figure}[t]
     \centering
     \subfloat[nug5\label{fig:best_feasibility_qap_nug5}]{
         \includegraphics[width=0.81\linewidth]{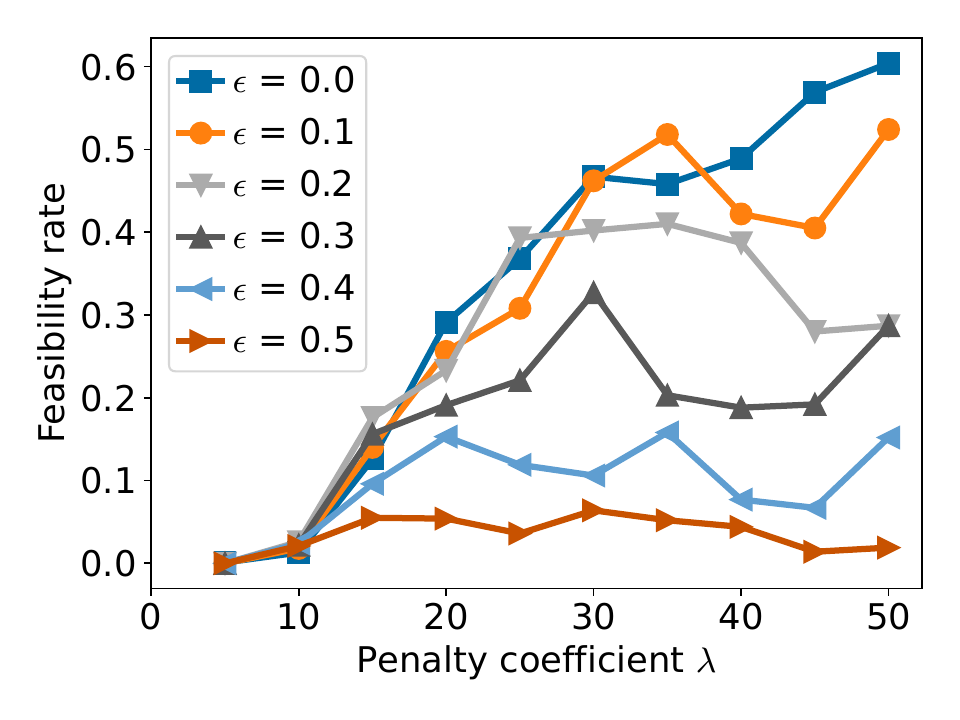}
     }
     \\ 
     \subfloat[tai5a\label{fig:best_feasibility_qap_tai5a}]{
         \includegraphics[width=0.81\linewidth]{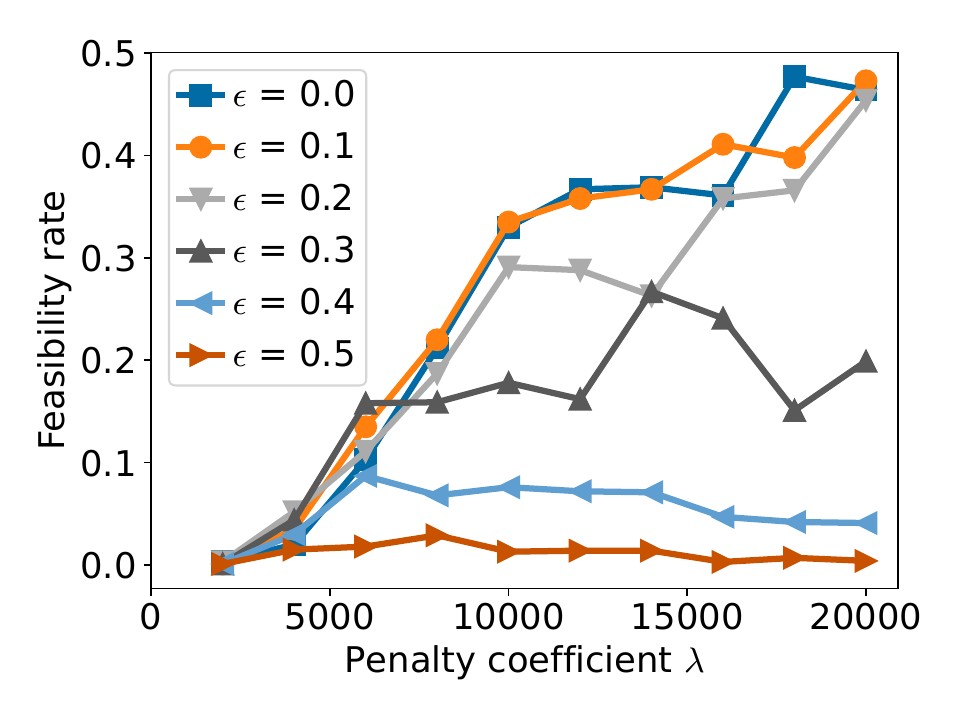}
     }
    \caption{
    Feasibility rate with \hl{optimal} chain strength on QAP \hl{instances}.
    }
    \label{fig:best_feasibility_qap}
\end{figure}

\begin{figure}[t]
     \centering
     \subfloat[nug5\label{fig:best_objective_qap_nug5}]{
         \includegraphics[width=0.81\linewidth]{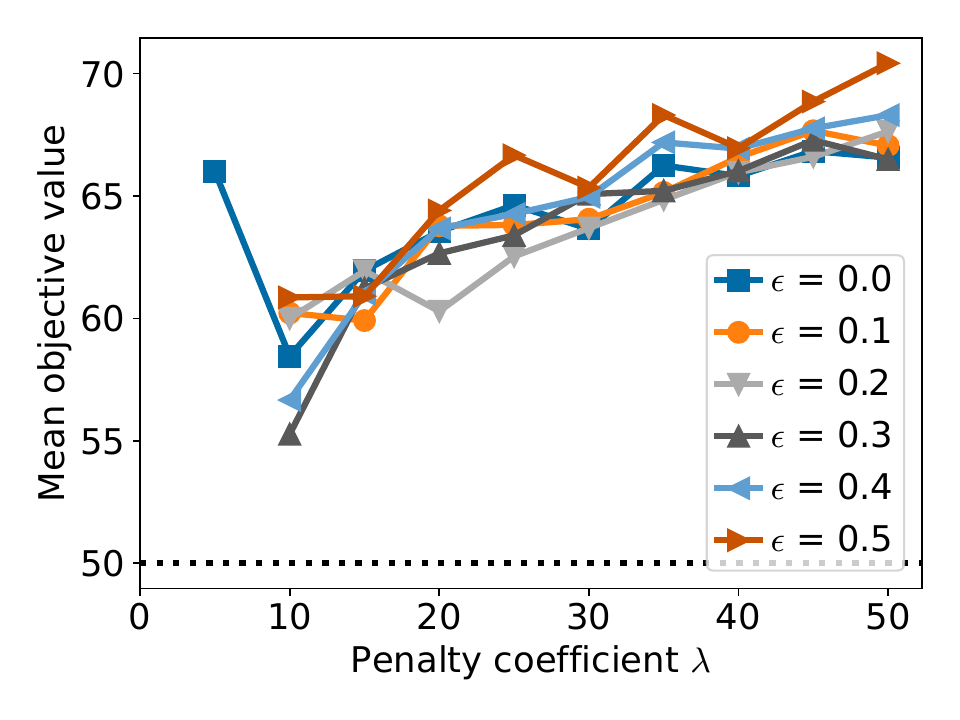}
     }
     \\ 
     \subfloat[tai5a\label{fig:best_objective_qap_tai5a}]{
         \includegraphics[width=0.81\linewidth]{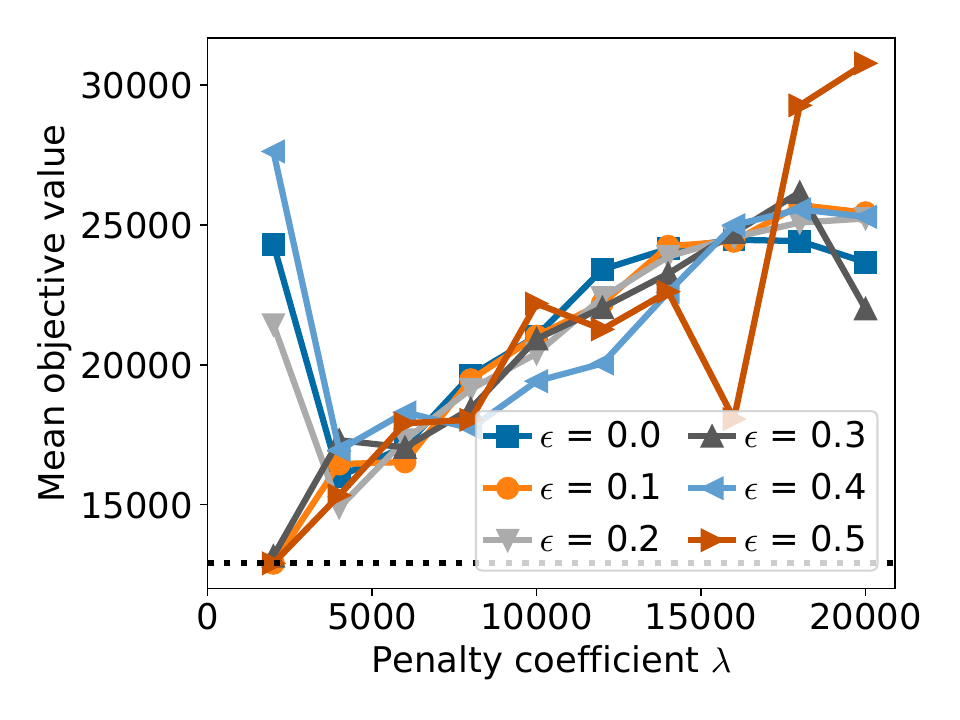}
     }
    \caption{
    Mean objective value with \hl{optimal} chain strength on QAP \hl{instances}. Dotted line represents optimum.
    }
    \label{fig:best_objective_qap}
\end{figure}

\hlbegin

We evaluate the ALM under minor-embedding.
First, we assess the scaling factor $s_{\tilde{h}}$ of the physical external field through minor-embedding search to ensure that the coupling strength 
limits the hardware precision.
Then, to evaluate the optimal chain strength and sample quality, we run quantum annealing to collect 100 solutions per configuration of the perturbation width $\epsilon$, penalty coefficient $\lambda$, and chain strength, for each of the 10 random seeds used in the minor-embedding search.
The optimal chain strength is determined based on average energy of the sampled solutions for each $(\epsilon, \lambda)$ pair.
To isolate the impact of the ALM from potential confounding effects, it is necessary to examine whether the perturbation $\epsilon$ significantly alters the optimal chain strength, as variations in the chain strength itself can influence sample quality.
Due to the large number of parameter combinations, we present only the results for the optimal chain strength here; the complete energy plots for all tested configurations are provided in Appendix~\ref{app:qap_energy_plots}.
Finally, we compare the sample quality (i.e., the feasibility rate and objective value) at the optimal chain strength with the SA results (where minor-embedding is not applied) to assess the reduction effect of the ALM in the presence of minor-embedding.

\hlx{We first present the comparison of the scaling factors $s_{\tilde{h}}$ and $s_J$.}
Fig.~\ref{fig:dr_qap} shows the ratio $s_{\tilde{h}}/s_J$ 
for tested perturbations and penalty coefficients.
We again confirm that $s_{\tilde{h}} \le s_J$ holds for all cases, which \hl{implies that the external fields do not limit the hardware precision for the QAP.
This result also supports the claim in Section~\ref{subsec:h_reduction_by_minor-embedding} that reducing external field coefficients is not necessary \hlx{on the QAP, even when the reduction method is applied}.}

\hlx{Fig.~\ref{fig:best_chain_qap} shows that} the optimal chain strength scales almost linearly with respect to the penalty coefficient.
This is as expected, since the penalty terms in 
Eq.~(\ref{eq:qap_penalized}) is dominant over the other components in terms of coefficient magnitude.
In contrast, the optimal chain strength remains largely consistent regardless of the perturbation width $\epsilon$.
This result confirms that the perturbation does not interfere with the chain strength setting, allowing us to focus exclusively on the benefits of penalty coefficient reduction when evaluating sample quality.
In other words, the reduction effect of the ALM on the physical coupling coefficients can be directly evaluated through its impact on the penalty coefficients, without adjusting for changes in the optimal chain strength.

To observe the reduction effect under minor-embedding, Fig.~\ref{fig:best_feasibility_qap} and Fig.~\ref{fig:best_objective_qap} present the feasibility rate and average objective value, respectively, with the chain strength fixed at its optimal value for each $(\epsilon, \lambda)$.
\hlend
Interestingly, the feasibility rate decreases almost monotonically with respect to the perturbation width $\epsilon$. In particular, the perturbation does not \hl{facilitate feasibility at} smaller penalty coefficients, \hl{which stands} in contrast to the SA results without minor-embedding.
\hl{
Similarly, regarding the objective value (Fig.~\ref{fig:best_objective_qap}), we do not observe any behavioral improvement due to perturbation, suggesting that the ALM does not enhance solution quality in this hardware setting.
Although the average objective value for $\epsilon>0$ occasionally outperforms the $\epsilon=0$ case at small $\lambda$, we consider these cases to be within the margin of error, since they involve only a few feasible solutions and the advantage is inconsistent across different $\epsilon>0$ values.}

The cause of \hl{the discrepancy between the two experiments} can be \hl{narrowed down to} two possibilities: \hl{the difference in} samplers (quantum vs simulated annealing) \hl{or the effect} of minor-embedding.
\hl{To distinguish between these,}
we conducted the same experiments using minor-embedding with SA and obtained results \hl{nearly identical to those from the quantum annealer}, see Appendix~\ref{app:qap_sa_pegasus}.
Therefore, \hl{it appears that the} minor-embedding \hl{itself alters} the behavior of samples under the perturbation.
\hl{While the underlying mechanism warrants further exploration, our results suggest that the ALM, or 
penalty perturbation, is not effective for reducing the penalty coefficient and may instead harm solution feasibility when used in conjunction with minor-embedding.}

\section{Conclusion and Future Perspectives}\label{sec:conclusion}

We verified whether existing coefficient-reduction methods are useful in compensating for the \hlx{limited} numerical precision of current quantum annealers.
This is the first \hlx{comprehensive} study to assess their \hlx{effectiveness under minor-embedding, which is a practical requirement for solving problems on actual hardware.}
Three existing methods are thoroughly evaluated by tracking the change of coefficients in the Hamiltonian due to minor-embedding, using benchmark instances of the quadratic unconstrained binary optimization (QUBO) problem, multi-dimensional knapsack problem (MKP), and quadratic assignment problem (QAP).
As a result, the interaction-extension method (IEM) successfully reduced the coefficients of the minor-embedded Hamiltonian and improved the sample quality of quantum annealing on the QUBO instances.
The bounded-coefficient encoding (BCE) of integer variables worked ideally on the toy problem, but the effect was limited when tested in a more practical setting using the MKP.
It is necessary to identify and resolve the cause of this gap to utilize the BCE effectively in current quantum annealers.
The verification using the QAP showed that the augmented Lagrangian method (ALM) \hlx{exhibits} the expected coefficient-reduction effect 
\hlx{in the absence of} minor-embedding, while this effect \hlx{disappears} when minor-embedding is applied, and in fact, the sample quality \hlx{may deteriorate}.
Further investigation into the reason of the significant decrease in the feasibility rate would lead to a better understanding of the effect of minor-embedding on quantum annealing.
These results do not imply that the IEM is the best reduction method since these methods have different applicability and properties from each other.
Rather, it would be important to explore how to use or improve these methods on the basis of the analysis on the gap between the expected and actual effects.

\hlx{Beyond the specific methods evaluated in this study, our methodology is inherently applicable to a broader class of reduction strategies, including those that may not strictly preserve the ground state. While the current work focuses on ground-state-preserving methods, the proposed framework can be utilized to quantify the practical utility of non-ground-state-preserving heuristics that may emerge in the future. This is particularly relevant in cases where the benefits of a reduced dynamic range might justify a potential loss of theoretical equivalence. Such assessments would provide a foundation for exploring coefficient-reduction approaches that prioritize hardware-limited performance over strict adherence to the original Hamiltonian.}

There are several possible directions of future study.
The first is to enhance the coefficient-reduction methods.
Our findings in this paper would be helpful to develop new, more effective reduction methods.
In particular, we found in the experiments that it is not necessary to apply the coefficient reduction to the external field in most cases.
It would be promising to consider approaches to reduce the coupling coefficients at the risk of increasing the external field coefficients, as in the ALM.
It is also interesting to explore the combination of quantum error correction (QAC)~\cite{pudenz2014error} and coefficient-reduction approaches.
It possibly enables us to adjust the trade-off between the hardware cost, e.g., the number of qubits, and sample quality, although this would be an even more complicated task when taking minor-embedding into account.
There is an existing paper~\cite{pelofske20244} related to this approach.
Another future direction is to develop a minor-embedding search strategy that takes the numerical precision into account.
One way is to assign a long chain to an external field term with a large coefficient to reduce it effectively.
\hlx{Another approach involves designing minor-embedding algorithms that minimize the required chain strength, which presents a more challenging task.}

\appendices

\section{Estimation of Scaling Factors on Large Instances in MQLIB}\label{app:physical_h_scaling_estimation}

In the experiment in Section~\ref{subsec:h_reduction_by_minor-embedding}, for a large instance which is not minor-embeddable to the hardware graph $P_{15}$, we consider a larger Pegasus graph $P_m$ to which the instance can be embedded.
Specifically, any graph having $n$ nodes can be embedded to $P_m$ with $m \ge \frac{n}{12}+1$ using clique-embedding~\cite{Dwaveadvantage}.
We choose the smallest $m$ satisfying the condition.
For the clique embedding, the chain length is $m$ or $m+1$.
Therefore, the scaling factor $s_{\tilde h}$ of the physical external fields is at most $s_{h}/m$, where $s_{h}$ is the scaling factor of the logical external fields.
We used this formula to estimate $s_{\tilde h}$ for non-embeddable instances.

\section{Preliminary Experiments on MKP}\label{app:mkp_preliminary}

\begin{figure}[t]
     \centering
     \subfloat[weing1\label{fig:tuning_objective_mkp_weing1}]{
         \includegraphics[width=0.81\linewidth]{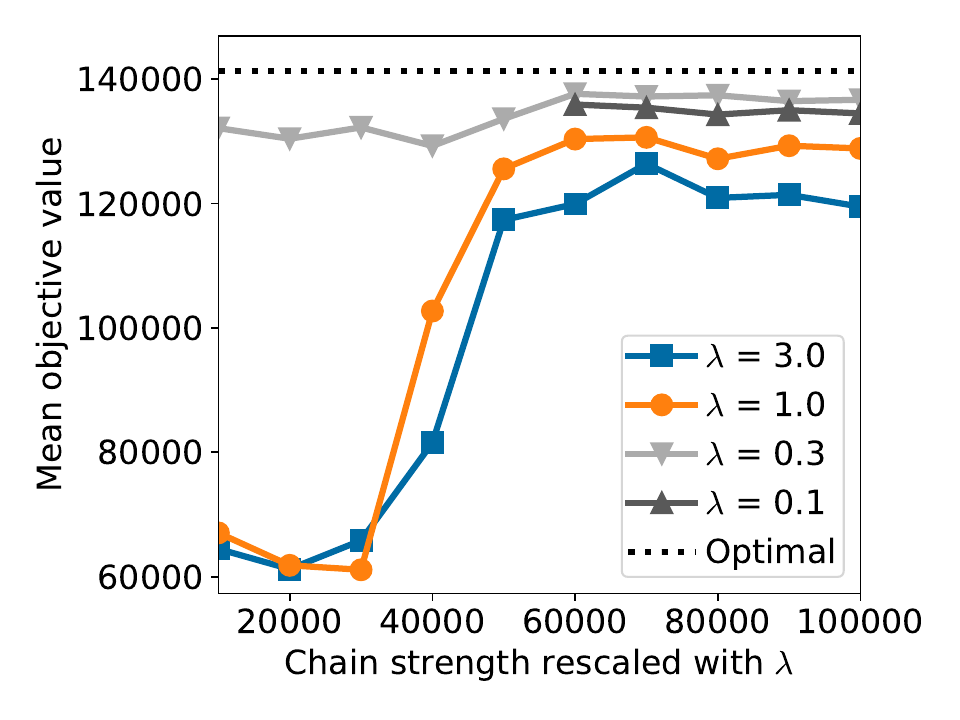}
     }
     \\ 
     \subfloat[weish06\label{fig:tuning_objective_mkp_weish06}]{
         \includegraphics[width=0.81\linewidth]{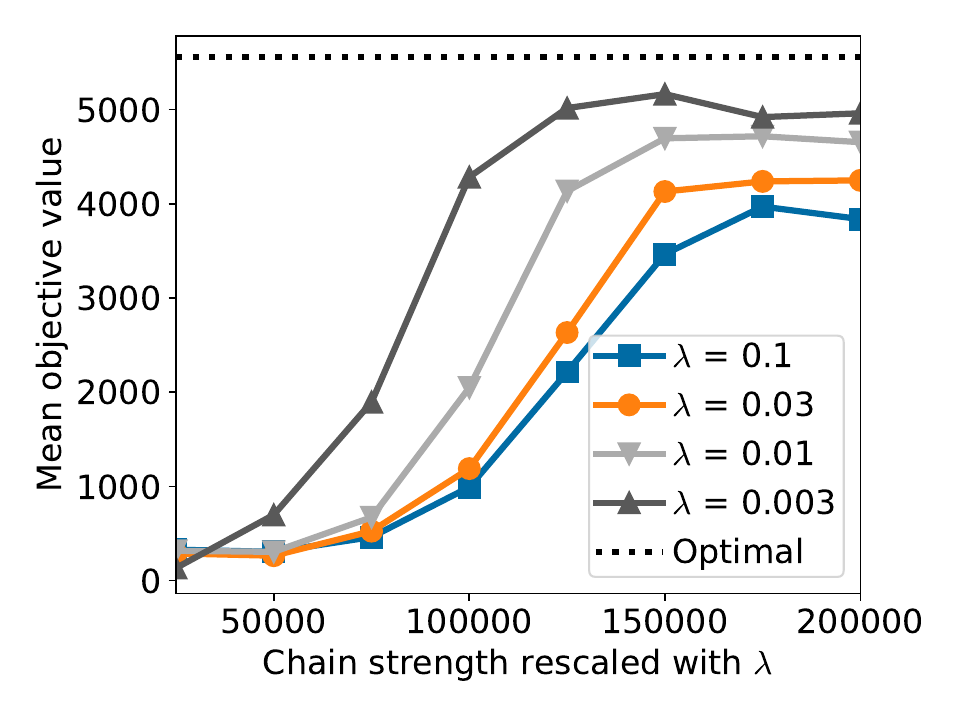}
     }
    \caption{
    Mean objective value on MKP \hl{instances} for various penalty coefficients $\lambda$.
    }
    \label{fig:tuning_objective_mkp}
\end{figure}

\begin{figure}[t]
     \centering
     \subfloat[weing1\label{fig:tuning_feasibility_mkp_weing1}]{
         \includegraphics[width=0.81\linewidth]{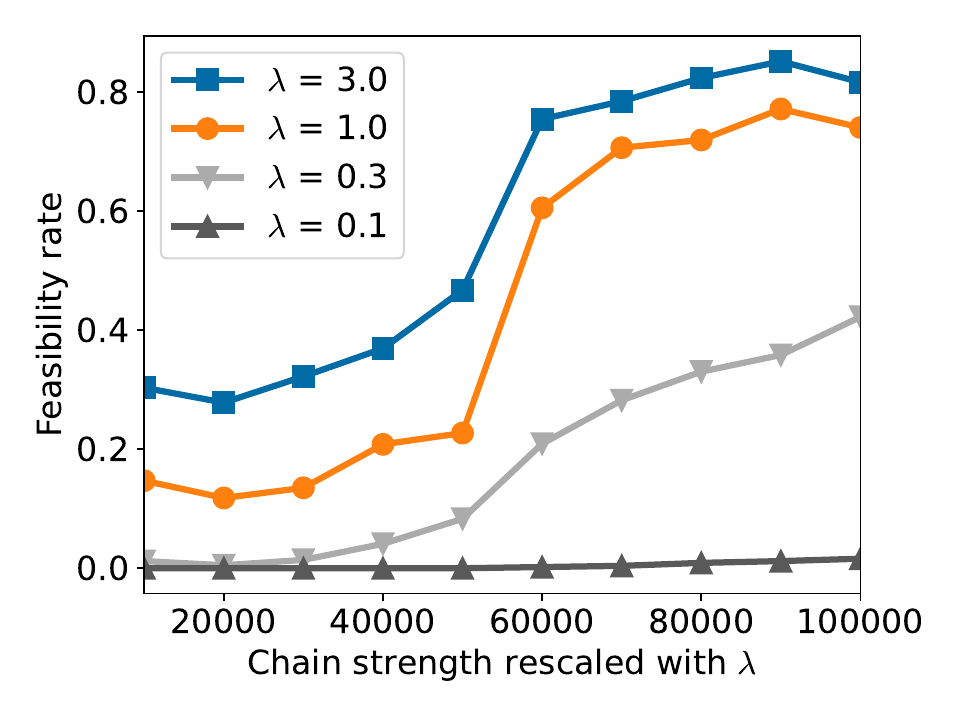}
     }
     \\ 
     \subfloat[weish06\label{fig:tuning_feasibility_mkp_weish06}]{
         \includegraphics[width=0.81\linewidth]{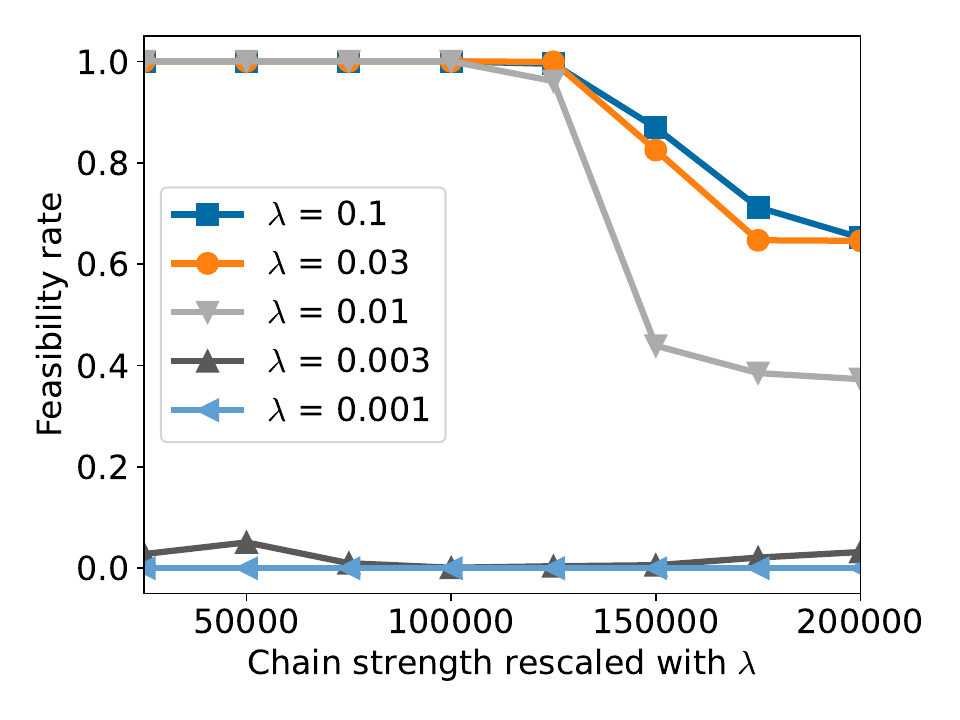}
     }
    \caption{
    Feasibility rate on MKP \hl{instances} for various penalty coefficients $\lambda$.
    }
    \label{fig:tuning_feasibility_mkp}
\end{figure}

\begin{figure}[t]
     \centering
     \subfloat[weing1\label{fig:dr_vs_bce_mkp_weing1}]{
         \includegraphics[width=0.81\linewidth]{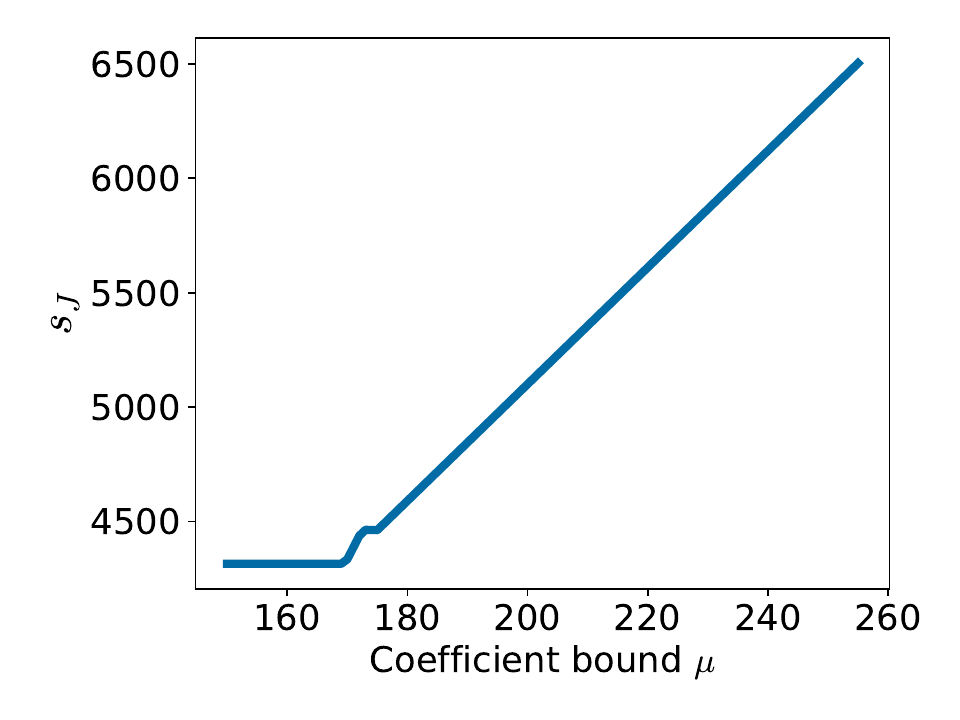}
     }
     \\ 
     \subfloat[weish06\label{fig:dr_vs_bce_mkp_weish06}]{
         \includegraphics[width=0.81\linewidth]{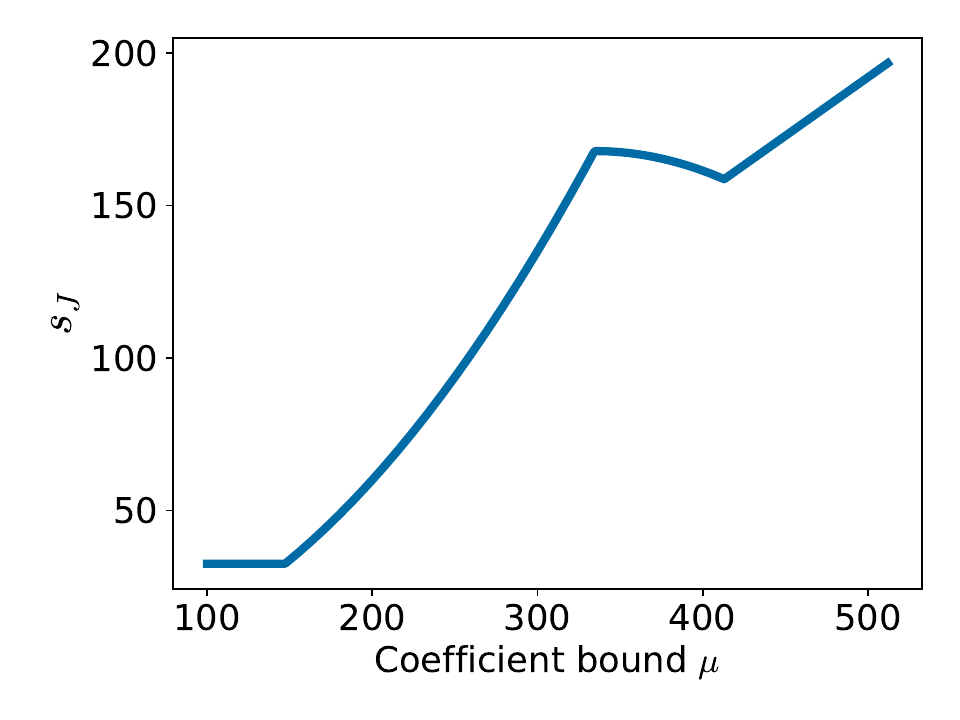}
     }
    \caption{
    Scaling factor $s_J$ of logical couplings \hl{on MKP instances}.
    }
    \label{fig:dr_vs_bce_mkp}
\end{figure}

We conducted preliminary experiments for \hl{Exp.~2 in Section~\ref{subsec:experiment_bce}} 
to decide the value of the penalty coefficient $\lambda$ and the range of the parameter $\mu$ in the BCE.

We chose the 
\hl{optimal} value of $\lambda$ in Eq.~(\ref{eq:mkp_penalized}) as follows.
We fixed $\mu$ to the maximum value, which corresponds to the binary \hlx{encoding} of $z_i$, and ran minor-embedding search with 10 random seeds.
We set the initial value $\lambda=10$ and repeated sampling 100 solutions for each embedding and chain and decreasing $\lambda$ by a factor of about $3$.
The procedure was stopped if no feasible solution is obtained.
\hl{Based on the results in Fig.~\ref{fig:tuning_objective_mkp} and Fig.~\ref{fig:tuning_feasibility_mkp}, the $\lambda$ values that yielded the best objective value were identified as 0.3 for weing1 and 0.003 for weish06.}

As for the range of $\mu$, we computed the scaling factor $s_J$ of logical couplings, varying the value of $\mu$.
The computed $s_J$ values are shown in Fig.~\ref{fig:dr_vs_bce_mkp}.
Since sufficiently large $\mu$ gives integer encoding equivalent to the binary \hlx{encoding}, the upper bound of $\mu$ is set to the smallest value resulting in the binary \hlx{encoding}.
On the other hand, when $\mu$ is decreased, $s_J$ reaches to a lower bound at some point and cannot be reduced further.
The point is given as $\mu=169$ on weing1 and $\mu=147$ on weish06.
This lower bound of $s_J$ is given by the maximum value of $\sum_{i=1}^m w_{ij}w_{ij'}$ for $j,j'=1,\ldots,n$, the coefficient appearing in the expansion of the QUBO objective function Eq.~(\ref{eq:mkp_penalized}).
Note that $s_J$ might increase even when $\mu$ is decreased, as shown in Fig.~\ref{fig:dr_mkp_weish06}.
This occurs only when $m$ in Eq.~(\ref{eq:bce_notation}) equals to 1.
In the validation experiments in Section~\ref{sec:experiments}, we chose a set of values of $\mu$ so that $s_J$ behaves monotonically with respect to $\mu$.

\section{Energy plots on QAP}\label{app:qap_energy_plots}

\begin{figure*}[t]
     \centering
     \subfloat[nug5\label{fig:energy_qap_nug5}]{
         \includegraphics[width=0.825\linewidth]{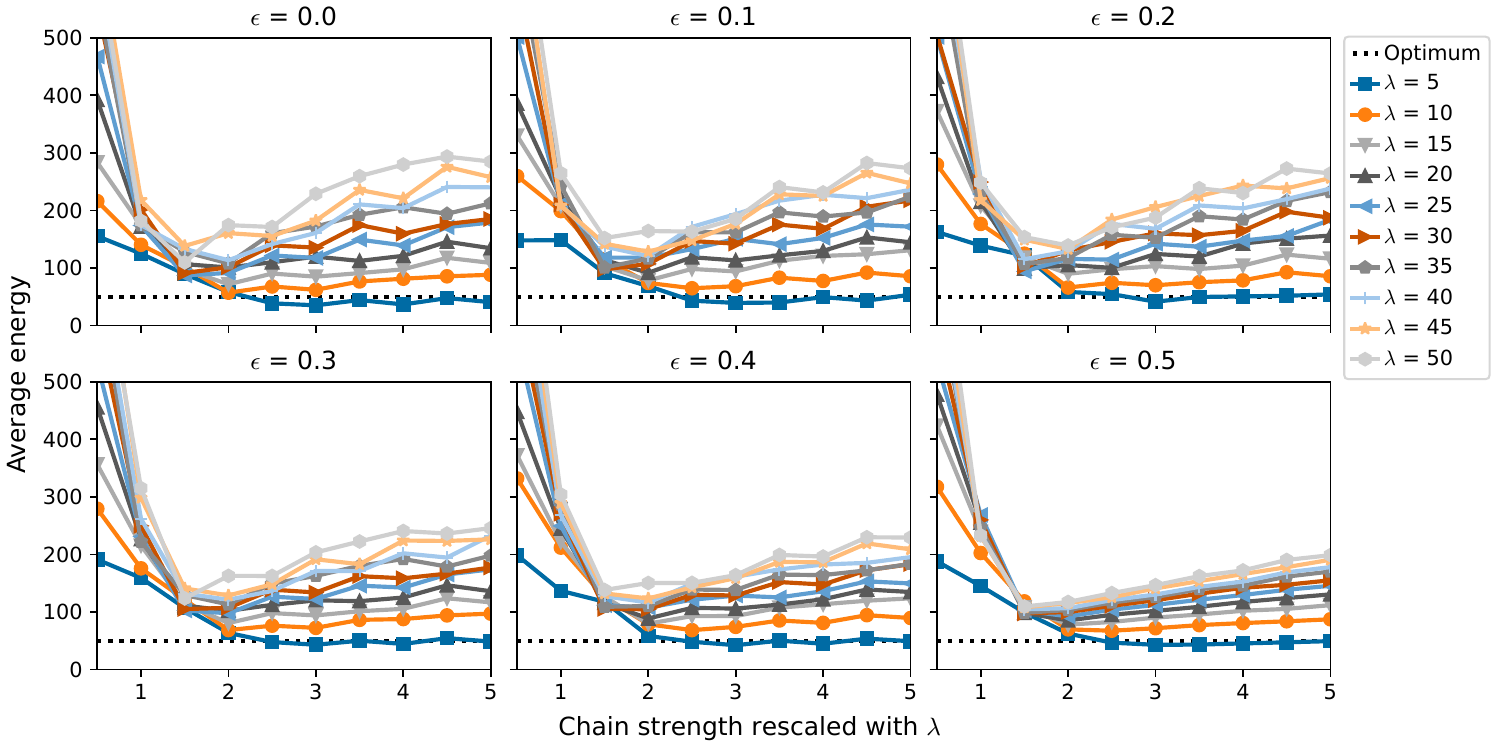}
     }
     \\ 
     \subfloat[tai5a\label{fig:energy_qap_tai5a}]{
         \includegraphics[width=0.825\linewidth]{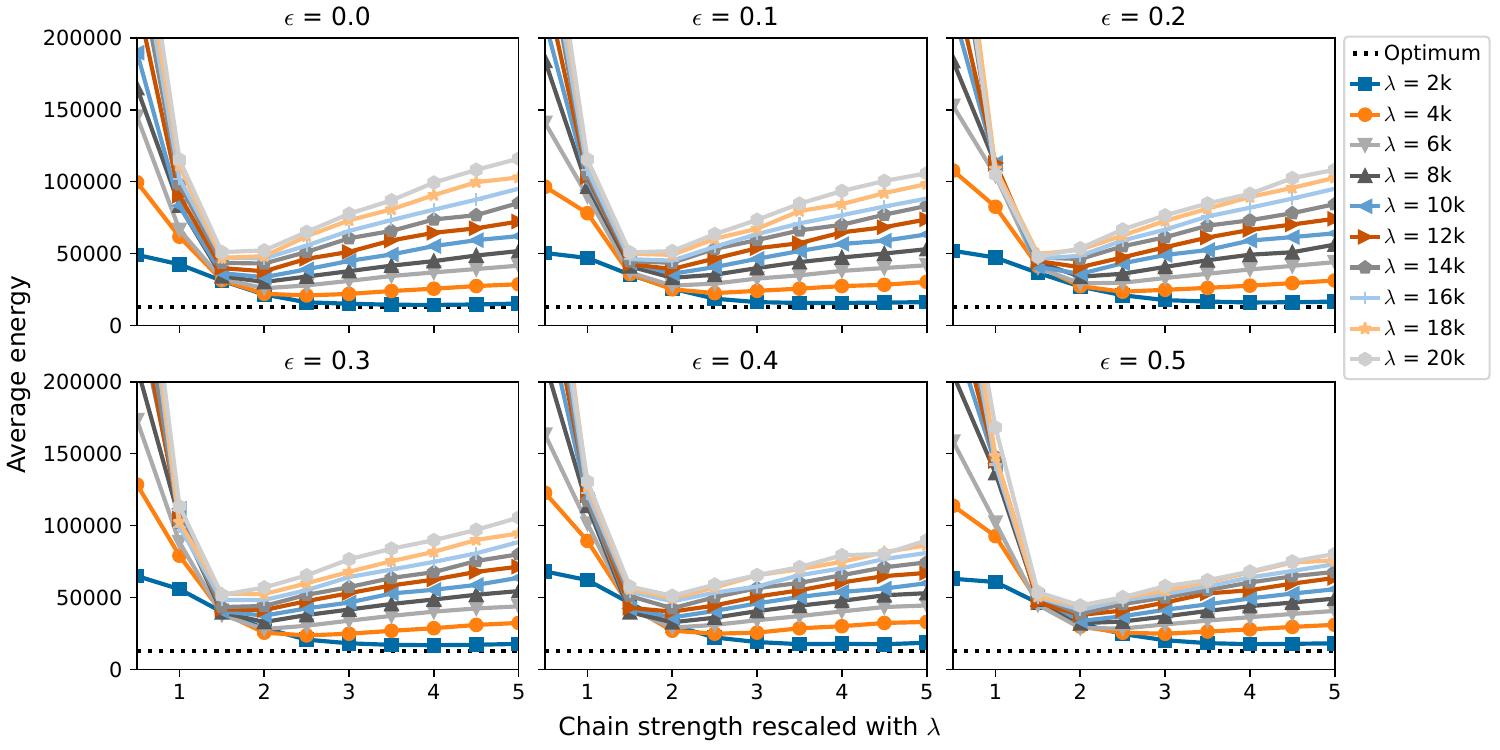}
     }
    \caption{
    Average energy of samples from the quantum annealer on the QAP \hl{instances}. It can be lower than the optimum objective value of the instance if $\lambda$ is small since samples can achieve lower energy by violating the constraint.
    }
    \label{fig:energy_qap}
\end{figure*}

We describe the result detail of experiments on QAP with quantum annealing in Section~\ref{subsec:experiment_augmented_lagrangian}.
Since we sample solutions varying the penalty coefficient $\lambda$, it is difficult to use the feasibility rate or the objective value, which are in the trade-off relation, for determining the \hl{optimal} chain strength.
Therefore, the \hl{optimal} chain strength in Fig.~\ref{fig:best_chain_qap} is computed based only on the average energy of samples.
The whole energy plots are shown in Fig.~\ref{fig:energy_qap}.
We observe that the behavior and location of energy peak are not much changed depending on the perturbation width $\epsilon$ also from the figure.

\section{Validation of ALM Using SA on minor-embedded Hamiltonian}\label{app:qap_sa_pegasus}

\begin{figure}[t]
     \centering
     \subfloat[nug5\label{fig:best_chain_qap_nug5_sa}]{
         \includegraphics[width=0.81\linewidth]{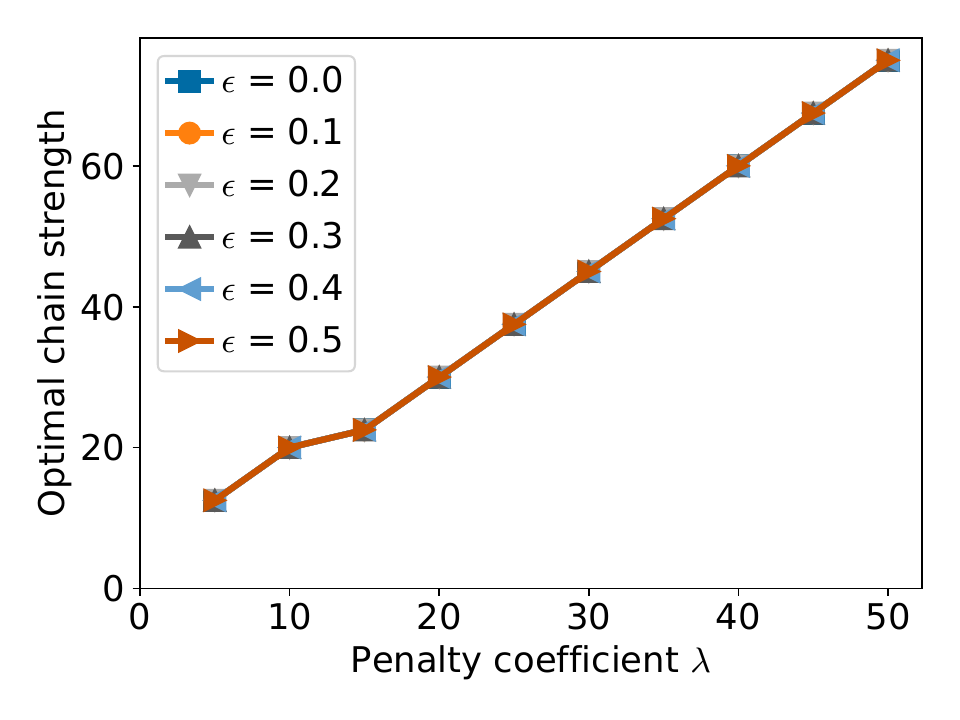}
     }
     \\ 
     \subfloat[tai5a\label{fig:best_chain_qap_tai5a_sa}]{
         \includegraphics[width=0.81\linewidth]{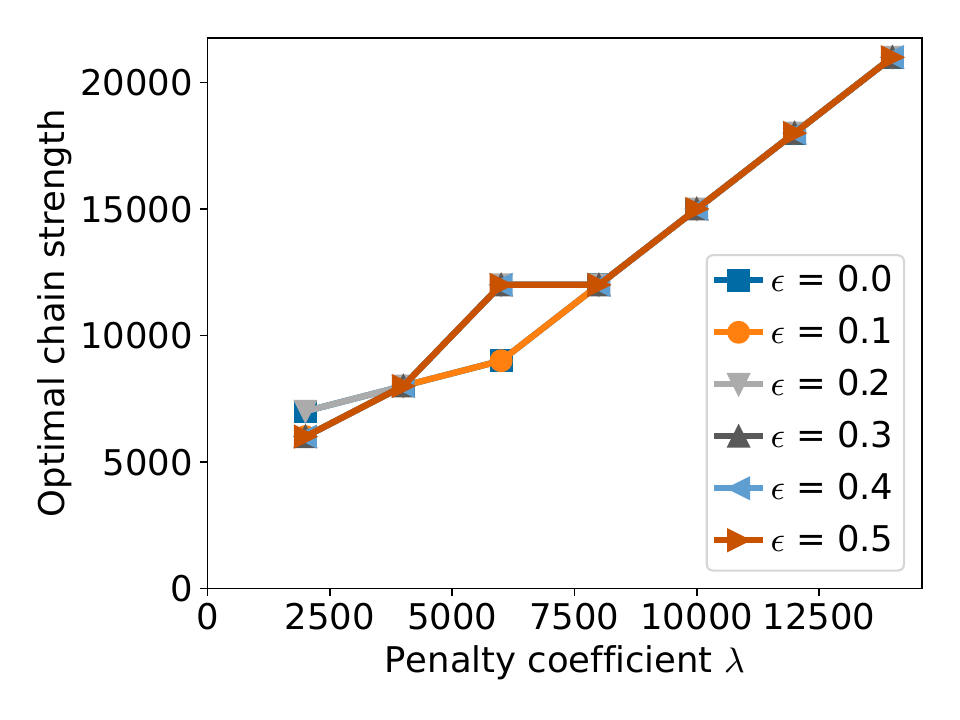}
     }
    \caption{
    \hl{Optimal} chain strength on QAP \hl{instances using} SA.
    }
    \label{fig:best_chain_qap_sa}
\end{figure}

\begin{figure}[t]
     \centering
     \subfloat[nug5\label{fig:best_feasibility_qap_nug5_sa}]{
         \includegraphics[width=0.81\linewidth]{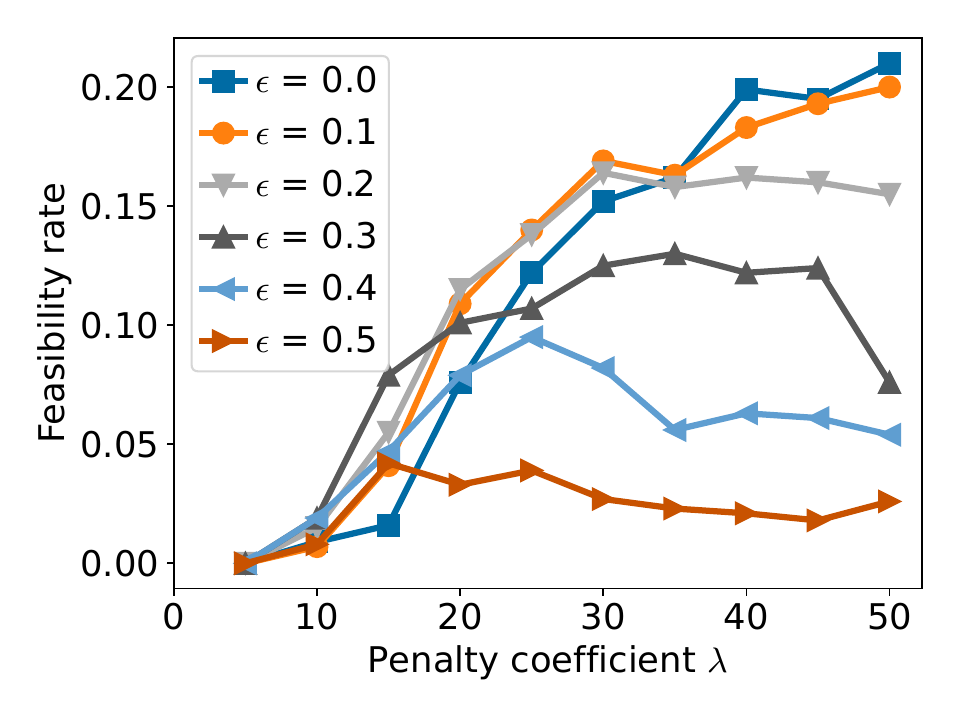}
     }
     \\ 
     \subfloat[tai5a\label{fig:best_feasibility_qap_tai5a_sa}]{
         \includegraphics[width=0.81\linewidth]{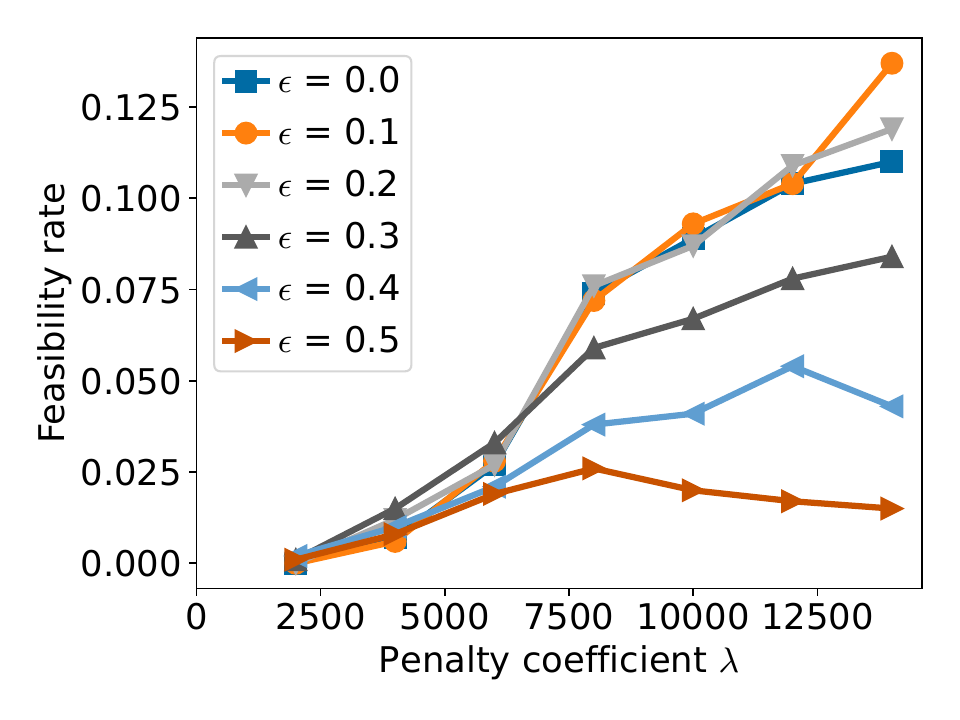}
     }
    \caption{
    Feasibility rate with \hl{optimal} chain strength on QAP \hl{instances using} SA.
    }
    \label{fig:best_feasibility_qap_sa}
\end{figure}

\begin{figure}[t]
     \centering
     \subfloat[nug5\label{fig:best_objective_qap_nug5_sa}]{
         \includegraphics[width=0.81\linewidth]{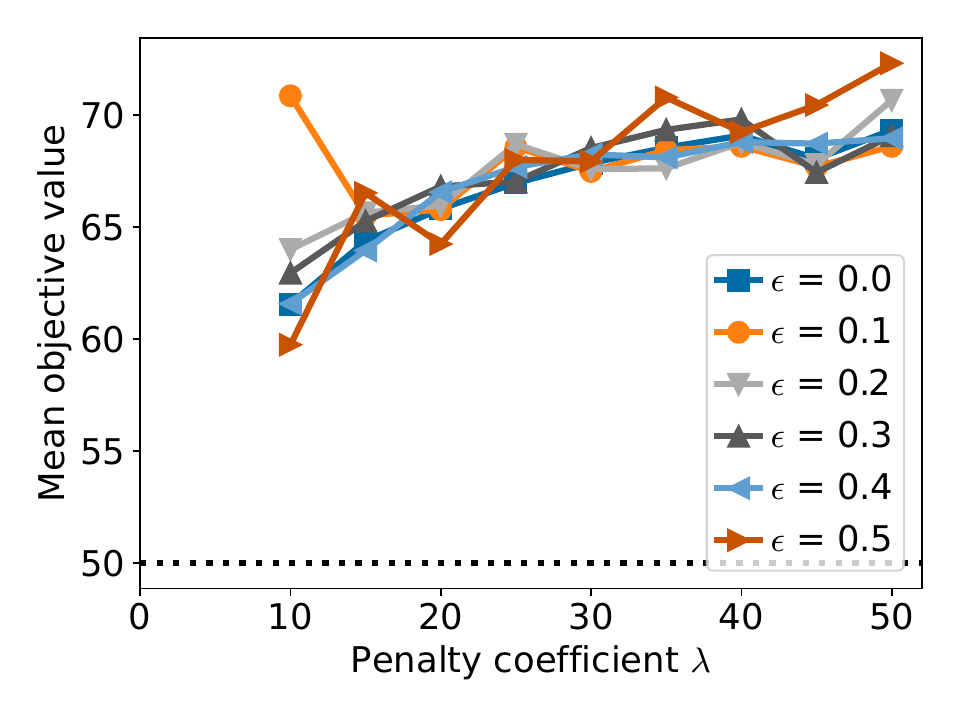}
     }
     \\ 
     \subfloat[tai5a\label{fig:best_objective_qap_tai5a_sa}]{
         \includegraphics[width=0.81\linewidth]{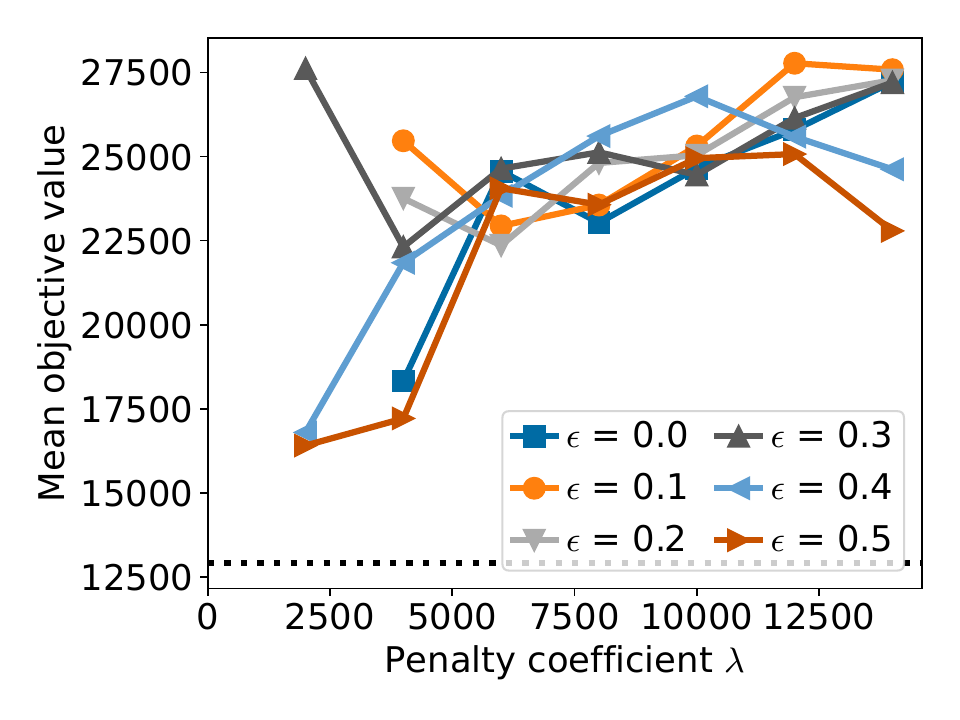}
     }
    \caption{
    Mean objective value with \hl{optimal} chain strength on QAP \hl{instances using} SA. Dotted line represents optimum.
    }
    \label{fig:best_objective_qap_sa}
\end{figure}

The experimental results of SA on the QAP with minor-embedding are provided in this \hl{appendix}.
We adopted the same experimental setup as the quantum annealer except that we apply SA instead of the quantum annealer after minor-embedding.
The \hl{optimal} chain strength, feasibility rate, and objective value are shown in Fig.~\ref{fig:best_chain_qap_sa}, \hl{Fig.~\ref{fig:best_feasibility_qap_sa}}, and \hl{Fig.~\ref{fig:best_objective_qap_sa}}, respectively.
The overall trend is quite similar to the result of the quantum annealer in Section~\ref{subsec:experiment_augmented_lagrangian}.
Therefore, we conclude that the gap between the result of SA without minor-embedding and that of the quantum annealer is caused by minor-embedding.

\bibliographystyle{IEEEtran}
\bibliography{ref}

\begin{IEEEbiography}[{\includegraphics[width=1in,height=1.25in,clip,keepaspectratio]{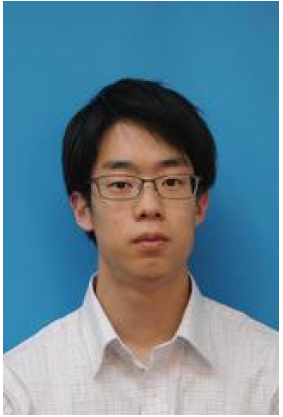}}]{Kentaro Ohno}
is a Ph.~D. student at Waseda University.
He received the B.~Sci. and M.~Sci. degrees in mathematics from the University of Tokyo in 2017 and 2019, respectively. From 2019 to 2024, he was a research engineer at NTT, Japan. He is currently studying combinatorial optimization using Ising machines.
\end{IEEEbiography}

\begin{IEEEbiography}[{\includegraphics[width=1in,height=1.25in,clip,keepaspectratio]{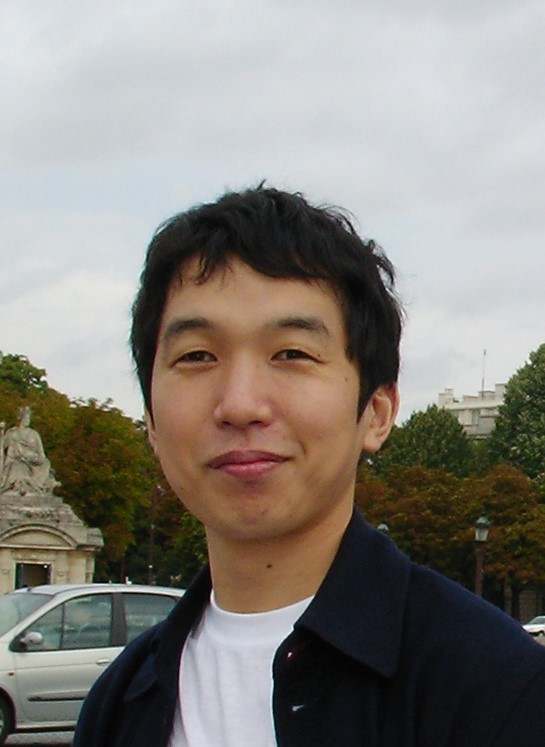}}]{Nozomu Togawa}
received the B.~Eng., M.~Eng., and Dr.~Eng. degrees from Waseda University in 1992, 1994, and 1997, respectively, all in electrical engineering. He is presently a professor in the Department of Computer Science and Communications Engineering, Waseda University. His research interests are quantum computation and integrated system design. He is a member of ACM, IEICE, and IPSJ.
\end{IEEEbiography}

\EOD

\end{document}